\documentclass[preprint2]{aastex}

\usepackage[caption=false]{subfig}
\usepackage{graphicx}
\usepackage{epstopdf}
\usepackage{amsmath}
\usepackage{longtable}
\usepackage{lscape}
\usepackage{natbib}
\usepackage{url}
\usepackage{floatfig}
\usepackage{color}
\usepackage[]{units}
\usepackage[]{nth}

\shorttitle{Flare Variability of Red-Dwarf Stars in M37}
\shortauthors{Chang et al.}

\begin{document}
\captionsetup[subfigure]{labelformat=empty}
\title{Photometric Study on Stellar Magnetic Activity: I. Flare Variability of Red Dwarf Stars in the Open Cluster M37}
\author{S.-W. Chang\altaffilmark{1}, Y.-I. Byun\altaffilmark{2}, and J. D. Hartman\altaffilmark{3}}

\altaffiltext{1}{Institute of Earth$\cdot$Atmosphere$\cdot$Astronomy, Yonsei University, Seoul 120-749, South Korea; seowony@galaxy.yonsei.ac.kr}
\altaffiltext{2}{Department of Astronomy and University Observatory, Yonsei University, Seoul 120-749, South Korea; ybyun@yonsei.ac.kr}
\altaffiltext{3}{Department of Astrophysical Sciences, Princeton University, Princeton, NJ 08544, USA}

\begin{abstract}
Based on one-month long MMT time-series observations of the open cluster M37, we monitored light variations of nearly 2500 red dwarfs and successfully identified 420 flare events from 312 cluster M dwarf stars.  For each flare light curve, we derived observational and physical parameters, such as flare shape, peak amplitude, duration, energy, and peak luminosity.  We show that cool stars produce serendipitous flares energetic enough to be observed in the $r$-band, and their temporal and peak characteristics are almost the same as those in traditional $U$-band observations.  We also found many large-amplitude flares with inferred $\Delta u > 6$ mag in the cluster sample which had been rarely reported in previous ground-based observations.  Following the ergodic hypothesis, we investigate in detail statistical properties of flare parameters over a range of energy ($E_{r}$ $\simeq$ $10^{31}$--$10^{34}$ erg).  As expected, there are no statistical differences in the distributions of flare timescales, energies, and frequencies among stars of the same age and mass group.  We note that our sample tend to have longer rise and decay timescales compared to those seen in field flare stars of the same spectral type and be more energetic.  Flare frequency distributions follow power-law distributions with slopes $\beta \sim0.62$--1.21 for all flare stars and $\beta \sim0.52$--0.97 for stars with membership information ($P_{mem} \geq 0.2$). These are in general agreement with previous works on flare statistics of young open clusters and nearby field stars.  Our results give further support to the classical age-activity relations.
\end{abstract}

\section{Introduction}
Stellar flares are good observational tracers of magnetic activity in low-mass stars (especially M dwarfs), since the intense release of flare energy is necessarily related to magnetic fields that are generated by dynamo process as on the Sun.  Stellar flares are often thought of as being in some way analogous to solar flares because they follow the same universal correlations over many orders of magnitude in energy, peak luminosity, and total duration at almost all wavelengths (e.g., EUV, Hard/Soft X-ray, UV, white light, and radio emissions).  However, the parameter range of stellar flares is much broader than that for the Sun.  They can also release 10$^{6}$ times more energy in total and have substantially strong field strengths (e.g., \citealt{asc08, ben10, shi11}). 

Since the seminal work by \citet{lac76}, a number of studies have examined the flare properties and statistics in optical regime, such as frequencies, amplitudes, time-scales, and energies for a handful of active M dwarfs (e.g. \citealt{mof74,pet84,let97,ish91,haw91,dal10}).  One of the important findings is that flare frequency distribution can be approximated by a power-law in energy, indicating that less energetic flares tend to occur more frequently.  The power-law distribution indicates that the flare process exhibits self-similar, scale-invariant statistics within the observed energy range \citep{ben10}.  

This is not a special property limited to the flare stars having an exceptionally high flaring rate.  \citet{kow09} quantified the flaring properties of $\sim$50,000 M dwarfs using the low-cadence photometric light curves in SDSS Stripe 82.  Based on these much larger and less biased sample, they confirmed that the amplitude, luminosity, and flaring rate of the SDSS flares are consistent with those found from dedicated photometric monitoring campaigns.  Another new result is that even inactive stars (no H$\alpha$ emission in the quiescent spectrum) can exhibit flare variability, as well as active ones.  A more extensive analysis of magnetically inactive stars is given in \citet{hil11} for the first time, according to which inactive stars flare less frequently than the active stars as expected.  

This kind of statistical study is extended to M dwarf flares in both the red-optical and near-infrared (NIR) regimes.  Using the combination of 2MASS and SDSS multi-epoch database, \citet{dav12} showed that the signatures of flares are detectable even in the NIR passbands.  They found that the frequency of NIR flare detection is about two orders of magnitude lower than those detected in optical bands.

Kepler data provides a new opportunity to measure the properties of white-light flares on solar- and late-type stars \citep{wal11,bal12,mae12,shi13}.  Among these studies, \citet{wal11} focused on $\sim$23,000 cool dwarfs with K--M spectral types and showed that flare stars closer to the Galactic plane are statistically younger and more likely to be magnetically active.  This age-activity relation is in good agreement with spectroscopic studies of M dwarfs in SDSS (e.g., \citealt{wes08,wes11}).

Star clusters offer excellent opportunities to examine how stellar magnetic activity depends on age and rotation rate.  However, such studies are rare due to the large amount of telescope time required to properly sample stochastic flare events.  Table \ref{Tab1} summarizes the previous optical, UV, radio, and X-ray observations of flares in the region of young- and old-aged open clusters (30 Myr$\sim$4 Gyr).  Most of flare samples are too small to obtain reliable statistics, except for the long-term optical monitoring data.  Moreover, only a few attempts were made to examine how many flare stars still exist in open clusters and how their flare properties change over the duration of the observations (\citealt{amb70,amb71,amb72,amb73,ger83}).

\begin{deluxetable}{clcccrrrc}
\tabletypesize{\scriptsize}
\tablecaption{Previous observations of flares in the region of young- and old-aged open clusters \label{Tab1}}
\tablewidth{0pt}
\tablehead{
\colhead{} & \colhead{} & \colhead{Age} & \colhead{Distance} & \colhead{} & \colhead{$N_\mathrm{star}\tablenotemark{b}$} & \colhead{$N_\mathrm{flare}\tablenotemark{b}$} & $t_\mathrm{obs}\tablenotemark{b}$ & \colhead{} \\
\colhead{Wavelength} & \colhead{Name} & \colhead{(Myr)} & \colhead{(pc)} & \colhead{Method\tablenotemark{a}} & \colhead{(\#)} & \colhead{(\#)} & (hours) & \colhead{References}
}
\startdata\\
Optical\phn & $\alpha$ Persei (Mel 20)\phn & 71\phn & 187\phn & PG/CCD\phn & 7\phn & 7\phn & 187\phn & a,b\phn \\
 & Pleiades (M45)\phn & 135\phn & 138\phn & PG\phn & 564\phn & 1,635\phn & 3,250\phn & a,c\phn \\
 &               & & & CCD\phn & 1\phn & 1\phn & 132\phn & d\phn \\
 & Ptolemy's Cluster (M7)\phn & 299\phn & 301\phn & PG\phn & 6\phn & 6\phn & 28\phn & e\phn \\
 & Coma Berenices (Mel 111)\phn & 449\phn & 96\phn & PG\phn & 14\phn & 21\phn & 338\phn & a\phn \\
 & Praesepe (M44)\phn & 729\phn & 187\phn & PG\phn & 59\phn & 146\phn & 680\phn & a\phn \\
 & Hyades (Mel 25)\phn & 787\phn & 45\phn & PE\phn & 2\phn & \nodata & \nodata & f\phn \\\\
\hline\\
UV\phn & Pleiades\phn & 135\phn & 138\phn & NUV\phn & 7\phn & 4\phn & 21.4\tablenotemark{c}\phn & g\phn \\
   & Hyades\phn   & 787\phn & 45\phn  & FUV/NUV\phn & 6\phn & 3\phn & 8.7\tablenotemark{c}\phn & g\phn \\\\
\hline\\
Radio\phn & Pleiades\phn & 135\phn & 138\phn & VLA (1.4 GHz)\phn & 40\phn & \nodata & 3\phn & h\phn \\
      &          &  &  & VLA (8.42 GHz)\phn & 4\phn & 1\phn & 2.6\phn & i\phn \\
      & Hyades\phn & 787\phn & 45\phn & VLA (5 GHz)\phn & 9\phn & \nodata & 2\phn & j\phn \\
      &  &  &  & VLA (1.5 GHz)\phn & \nodata & \nodata & 14\phn & k\phn \\\\
\hline\\
X-ray\phn & NGC 2547\phn & 36\phn & 455\phn & XMM\phn & 108\phn & 7\phn & 13.7\phn & l \phn \\
      & Blanco 1\phn & 63\phn & 269\phn & XMM\phn & 33\phn & 7\phn & 13.9\phn & m\phn \\
      & $\alpha$ Persei\phn & 71\phn & 187\phn & ROSAT\phn & 71\phn & 3\phn & 6.3--6.9\phn & n\phn \\
      & NGC 2516\phn & 113\phn & 409\phn & XMM\phn & \nodata & 4\phn & 27.2\phn & o\phn \\
      &              &  & & Chandra\phn & 139\phn & 5\phn & 20.1\phn & p\phn \\      
      & Pleiades\phn & 135\phn & 138\phn & ROSAT\phn & 24\phn & 1\phn & 1.1\phn & q\phn \\
      &          &  &  & ROSAT\phn & 171\phn & 12\phn & 5.7--7.6\phn & r,s\phn \\
      &          &  &  & Chandra\phn & 18\phn & 11\phn & 17.2\phn & t\phn \\
      & Hyades\phn   & 787\phn & 45\phn & ROSAT\phn & 185\phn & \nodata & 0.05--0.16\phn & u\phn \\
      & NGC 752\phn  & 1,122\phn & 457\phn & Chandra/XMM\phn & 21/19\phn & 1/1\phn & 38.9/13.9\phn & v\phn \\
      & NGC 188\phn & 4,285\phn & 2,047\phn & XMM\phn & 6\phn & 1\phn & 11.4\phn & w\phn \\\\
\enddata
\tablecomments{We compiled a sample of flare stars in stellar clusters serendipitously detected at different wavelengths (optical, UV, radio, and X-rays between 1980 and 2013.  Not all flare stars are confirmed as cluster members.  The age and distance of each cluster were taken from the latest version of open cluster catalogs \citep{dia02,mer00}.  References are as follows: (a) \citet{tsv12}; (b) \citet{sem00}; (c) \citet{har82}; (d) \citet{mou11}; (e) \citet{jon91}; (f) \citet{pet89}; (g) \citet{bro09}; (h) \citet{bas88}; (i) \citet{lim95}; (j) \citet{cai89}; (k) \citet{whi93}; (l) \citet{jef06};  (m) \citet{pil05}; (n) \citet{pro96}; (o) \citet{ram03}; (p) \citet{wol04}; (q) \citet{sch93}; (r) \citet{sta94}; (s) \citet{gag95}; (t) \citet{dan03}; (u) \citet{ste95}; (v) \citet{gia08}; (w) \citet{gon05}.}
\tablenotetext{a}{Photographic plate observations: PG; Photoelectric observation: PE; CCD observations: CCD; {\sc GALEX} FUV/NUV imaging observations: FUV/NUV; Very Long Array radio observations: VLA; {\sc ROSAT} X-ray observations: ROSAT; {\sc XMM-Newton} X-ray observations: XMM; {\sc Chandra} X-ray observations: Chandra}
\tablenotetext{b}{$N_\mathrm{star}$ is the number of observed known (or candidate) flare stars; $N_\mathrm{flare}$ is the number of detected flare events; $t_\mathrm{obs}$ is the observing time.  Note that in the case of X-ray samples, $N_\mathrm{star}$ denotes the number of observed X-ray sources among the possible cluster members.}
\tablenotetext{c}{The Hyades was observed for a total of 4.44 hours and 4.23 hours respectively in both of the FUV and NUV imaging channels, while the Pleiades was observed solely in the NUV channel.}
\end{deluxetable}

We hereby investigate the statistical properties of flare and starspot-induced variabilities simultaneously; and seek to understand the relation between age, activity, and rotation in open cluster stars.  In this first paper, we present the flare properties and statistics for groups of stars with the same age and mass range in the open cluster M37.  Our second paper will deal with rotational properties of the same cluster stars.  This intermediate-age (550 Myr) cluster is well-suited target for detection of stellar flares, since a significant fraction of low-mass members are still magnetically active.
 
In Section 2, we briefly describe archival imaging data of the M37 taken by MMT 6.5m telescope and sample selection for flare searches.  We present the details of flare detection procedure in Section 3.  In Section 4, we derive the observational and physical parameters of individual flares.  Section 5 discuss the statistical properties of flare time-scales, energies, and frequencies.  Our conclusions follow in the last section.

\section{Data description}
We used the archival imaging data of the M37 taken by MMT 6.5m telescope (see \citealt{har08} for details).  This is from the survey designed to monitor the M37 field ($\square \simeq 24^{\prime}\times24^{\prime}$) for detection of possible exo-planets. The archive contains almost 5000 images obtained in $r\arcmin$-band over one-month period.  We refer the reader to \citet{cha15a} for a detailed description of our new photometric reduction and light curve production.  New light curves allow the analysis of brief transients such as flares.  The cadence of observations is between $\sim$80 to $\sim$150 seconds for over 90\% of the data.

\subsection{Sample selection}
We select the cluster M dwarf candidates by their photometric color and location on the color-magnitude diagram (CMD) of the M37 field.  It is known that the color range of $r - i > 0.53$ and $i - z > 0.3$ is occupied by the M and L dwarf classes \citep{wes05}.  Since these two colors are sensitive to changes in temperature for cool stars, it allows for proper separation of the stellar locus as a function of spectral type.  \citet{kow09} noted that the limiting $r - i$ color is reduced to 0.43 and the $i - z$ color to 0.23, after correcting for the Galactic reddening.  For cluster members, we correct the MMT $gri$ magnitudes for extinction by using the conversion relations in \citet{sto02}: $A_{g}/E(B-V)$ = 3.793, $A_{r}/E(B-V)$ = 2.751, and $A_{i}/E(B-V)$ = 2.086, respectively.  We adopted the cluster parameters from \citet{har08}: $E(B-V) = 0.227\pm0.038$, $(m-M)_{V} = 11.572\pm0.13$ for this procedure.

\begin{figure}[!t]
\centering
  \includegraphics[width=\linewidth, angle=0]{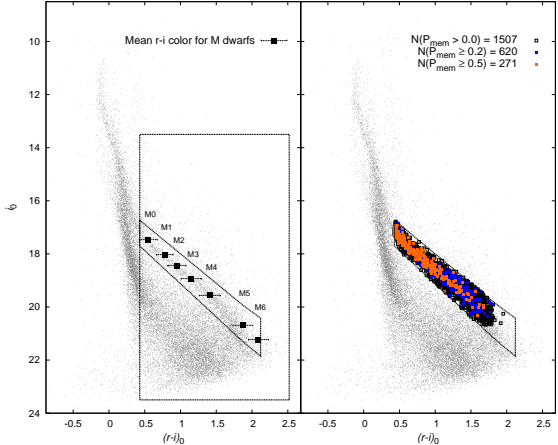}  
  \caption{De-reddened $r - i$ vs. $i$ color-magnitude diagram for sample selection. \emph{Left panel}: the outer dashed line represents a total of 15,873 stars that were initially selected by color of $(r - i)_{0} > 0.43$.  Among them, only 2459 cluster region stars are within the inner dashed box.  The mean $r - i$ colors for M dwarfs are indicated by squares and their 1-$\sigma$ error bars \citep{kow09}. \emph{Right panel}: the CMD positions of stars with cluster membership probabilities $P_{mem}$ are indicated by different colors, black for $P_{mem} \geq 0.0$, blue for $P_{mem} \geq 0.2$ and orage for $P_{mem} \geq 0.5$.}
  \label{Fig1}
\end{figure}

Using above relations, a total of 15,873 sources were initially selected by the de-reddened color of $(r - i)_{0} > 0.43$ (the outer dashed rectangular box in the left panel of Figure \ref{Fig1}).  We did not restrict the $i - z$ color because the $z$-band photometry of this subsample is often unreliable for faint sources.  In order to reduce cluster membership ambiguities, we divide the sample into the two groups: cluster stars and field stars according to their CMD locations.  We use the terms \lq\lq{}cluster sample\rq\rq{} and \lq\lq{}field sample\rq\rq{} while noting that the cluster sample is contaminated by field stars.  \citet{har09} statistically evaluated the contamination of photometrically selected cluster sample to be about 25--50\% as a function of magnitude along the main sequence, but it is less reliable at the faint end.  To improve the membership information, we used a membership probability ($P_{mem}$) based on the position of individual stars in the CMD and their radial distance from the cluster center (see \citealt{nun15} for details).  They considered stars with $P_{mem} \ge 0.2$ to be candidate cluster members since the effect of field star contamination will not be great on rotation-activity analysis of cluster stars.  As shown in the right panel of Figure \ref{Fig1}, many stars with $P_{mem} \ge 0.2$ (filled blue squares) are included in the inner dashed box (except for faint red stars).  In the following sections, we shall address the effect that field star contamination has on our results later.

\begin{figure*}[t]
\centering
  \includegraphics[width=0.95\linewidth, angle=0]{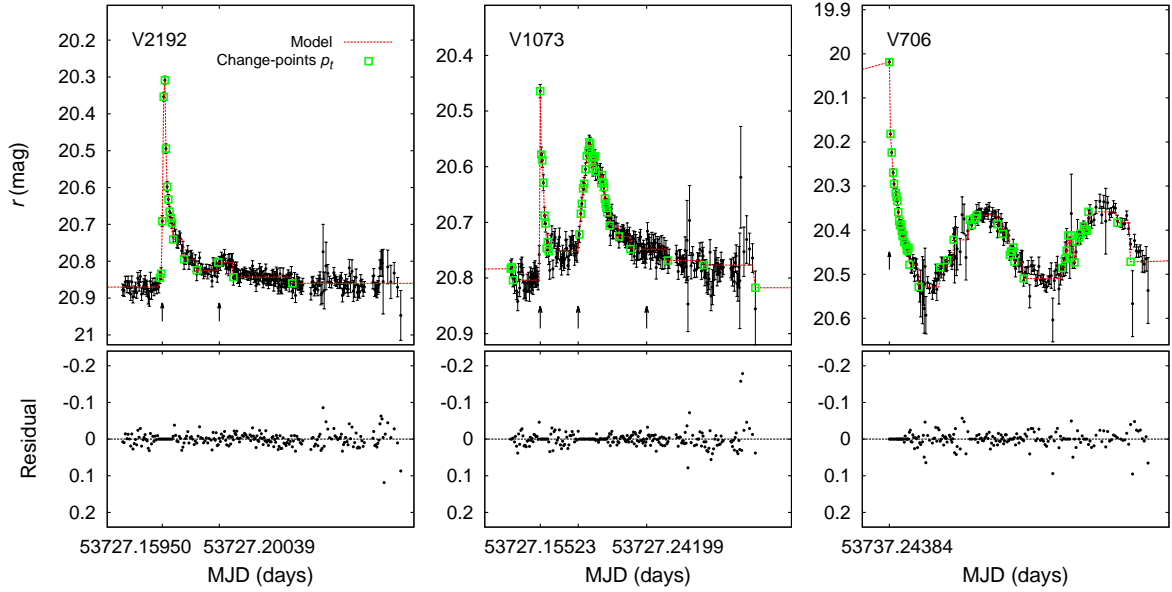}  
  \caption{Examples of flares detected by FINDflare algorithm.  As shown in the residual plots, the observed data are approximated by a piecewise constant model (red dashed lines).  The assumed start times of flare occurrence are indicated by the arrow.  Interestingly, our algorithm often detects secondary flares which occur during the decay of a much larger flare (V2192 and V1073).  The object IDs are taken from our new variable catalog of the M37 (see \citealt{cha15b}).}
  \label{Fig2}
\end{figure*}

The inner dashed box in the left panel of Figure \ref{Fig1} gives the sample selection criteria for the cluster sample.  This led to the identification of 2459 stars along the cluster sequence in the $i_{0}$ vs. $(r - i)_{0}$ CMD.  Based on the mean $r - i$ color for a given M dwarf spectral type (see Table 1 in \citealt{kow09}), each spectral type is indicated by the squares which lie along an extension of the cluster sequence.  Among them, only 620 stars have cluster membership probability with $P_{mem} \geq 0.2$.

\section{Variability analysis}

\subsection{Change-point analysis for flare detection (FINDflare)}
Flare-like features need to be detected from light curves without a priori knowledge of its shape and underlying brightness variations.  Our approach is very simple: we get a series of estimated {\em change-points} which correspond to the moments of apparent systematic changes in brightness.  The procedure is based on the statistics given by the cumulative sum (CUSUM) chart with bootstrap re-sampling, which is briefly described below  (see \citealt{cha15b} for details).

We have a set of discrete time-series data points \{$x_{1}$, $x_{2}$, $\cdots$, $x_{n}$\}, where $n$ is the time of occurrence of the $n$th data point.  The CUSUM values are given by:
\begin{equation}
S_{t} = \sum_{i=1}^{t} (x_{\mathrm{i}} - \bar{x})
\end{equation} for $t$ = $0, 1, \cdots, n$, where $S_{\mathrm{0}}$ = 0 and the mean $\bar{x}$.  The distribution of CUSUM values often has an inflection point at which the sign of the CUSUM slope changes, indicating that a significant deviations occur at this time.  This is used to determine whether a given interval of data should be kept as one ($\bar{x}_1$=$\bar{x}_2$=$\cdots$=$\bar{x}_n$=$\bar{x}$) or subdivided into two subintervals ($\bar{x}_1$=$\cdots$=$\bar{x}_{p}$$\neq$$\bar{x}_{p+1}$=$\bar{x}_n$).  To reduce the rate of false positives, we estimated the confidence level (c.l.) based on the bootstrap procedure.  Once sudden change has become apparent for a given region (c.l. $>$ 90\%), the location of change-point is initially determined as follows:
\begin{equation}
p_{t} = \operatorname*{arg\, max}_{t\in\left[0,n\right]} \left|S_{\mathrm{t}}\right|,
\end{equation} where $p_{t}$ denotes the last point before the change occurred.  The time-series data is split into two segments on each side of the change point, and the analysis is repeated for each segment until no more significant change point is detected.  All identified change-points thus define the segments (i.e., piecewise constant level sets), characterized by the start and end time of a given interval with its mean and variance. 

Lastly, we perform flare detection process for each segment.  For the $i$th measurement of a light curve, a simple selection criteria is given by the expressions:
\begin{mathletters}
\begin{equation}
x_\mathrm{i} - {\bar{x}_\mathrm{L}} < 0, 
\end{equation}
\begin{equation}
\frac{\left|x_\mathrm{i} - {\bar{x}_\mathrm{L}}\right|}{\sigma_\mathrm{L}} \ge N_\mathrm{1},
\end{equation}
\begin{equation}
\frac{\left|x_\mathrm{i} - {\bar{x}_\mathrm{L}} + w_\mathrm{i}\right|}{\sigma_\mathrm{L}} > N_\mathrm{2},
\end{equation}
\begin{equation}
ConM \ge N_\mathrm{3},
\end{equation} where the mean $\bar{x}_\mathrm{L}$ and deviation $\sigma_\mathrm{L}$ are the local statistics for a given segment, $w_\mathrm{i}$ is the photometric error at epoch $i$, and $ConM$ is the number of consecutive points which satisfy the equations (3a--3c).  The values of $N_{1,2,3}$ are taken to be at least larger than 3, 1, and 3, respectively.
\end{mathletters}

Figure \ref{Fig2} shows the combined result of the change-point analysis and identification of flare candidates in stars with and without underlying variability.  The light curves are well approximated by piecewise constant model with several discrete segments.  This model simplifies the task of detecting significant deviations from the mean level (i.e., statistical outlier).  Also, our algorithm can identify flare candidates, regardless of intrinsic variability.   

Many of flare candidates are short-duration ($3\le N_{3}\le5$) events.  Therefore their variability is suspected especially when such events are only partially observed, and found either at the end or beginning of available data. Our algorithm compares the light curves of flare candidates with those of nearby stars to ensure that these events are real.

\subsection{Removal of periodic variability}
Cool spots on the surface of stars are the dominant cause of periodic variability seen in some flare stars (e.g., V706 in Figure \ref{Fig2}).  Since these underlying variations can affect the extraction of flare parameters (see Section 4), we removed quasi-periodic patterns by a least-square harmonic fit to the data.  The model function is expressed as follows, which comprises a Fourier series truncated at harmonic $h$:
\begin{equation}
m_{h}(t) = \bar{m}_{0} + \sum_{h=1}^{4} A_{2h-1} \sin(B) + A_{2h} \cos(B); 
\end{equation} 
\begin{displaymath}
B = h\cdot2\pi t/P,
\end{displaymath} where $\bar{m}_{0}$ is the global mean for a whole range of data, $A_{2h}$ and $A_{2h-1}$ are the amplitudes obtained in each harmonic model, and $P$ is the period of the star.  The initial periods are taken from tables in our new catalog of variable stars in the field of M37 (see \citealt{cha15b}).  As a conservative approach, we have re-estimated the rotational period after rejecting outliers $>\pm3\sigma$, including flare-like features.  Then, the best-fit model is determined by comparing the difference of the reduced chi-squared values ($\Delta\chi_{\nu}^{2}$) before and after subtracting the harmonic models ($h$=1--4).  A total of 401 stars out of the 2495 cluster sample exhibit clear quasi-sinusoidal variations, which are removed from light curves to ensure proper characterization of flare events.

\begin{figure}[t]
\centering
    \includegraphics[width=\linewidth, angle=0]{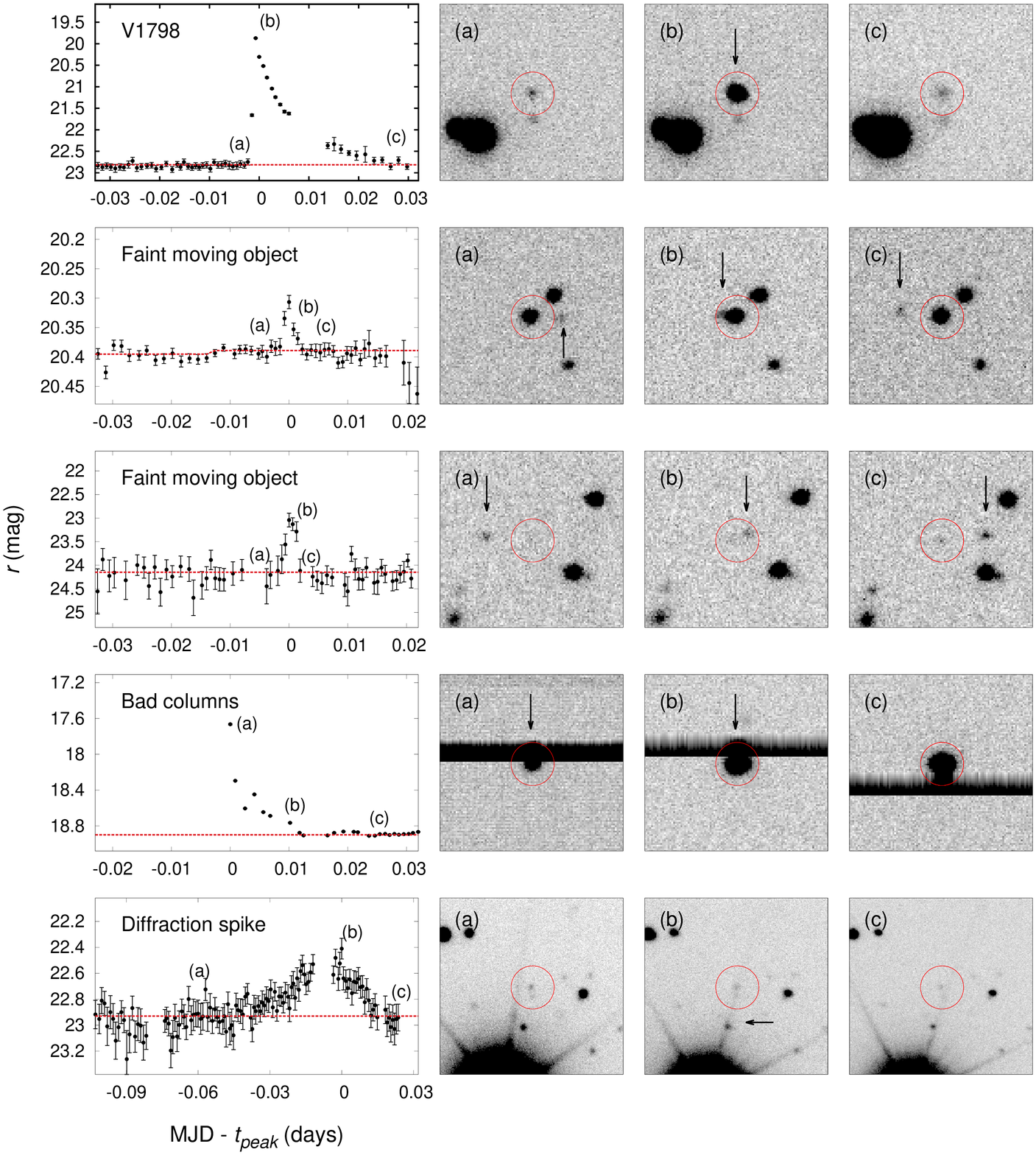}
    \caption{Example light curves of short, local flux enhancements detected by FINDflare algorithm.  From top to bottom, all except for the top panel (real flare event) are false positives due to contamination by moving objects, bad columns, and diffraction spikes around the bright stars.}
  \label{Fig3}
\end{figure}

\begin{figure}[t]
\centering
  \includegraphics[width=\linewidth, angle=0]{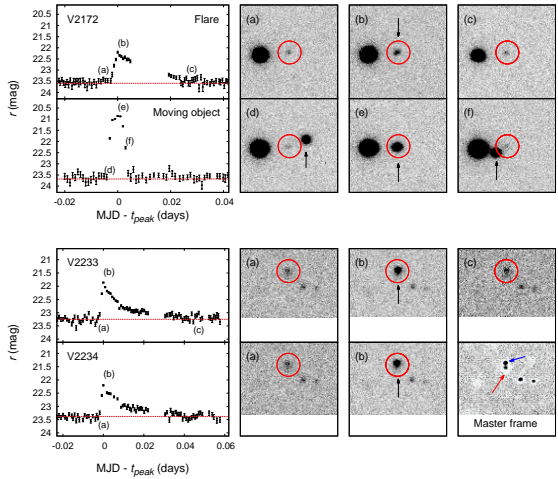}
  \caption{Same as Figure \ref{Fig3}, but for two exceptional cases.  In the top case, both real flare and false-positive events were observed on the same star. While in the bottom one, the light curve of one object shows flare-like variability, but this is actually a flare in a adjacent star.  From the combined deep image, i.e., master frame, we confirm a close companion with a separation of 0.79 arcseconds.}
  \label{Fig4}
\end{figure}

\begin{figure}[!t]
  \includegraphics[width=\linewidth, angle=0]{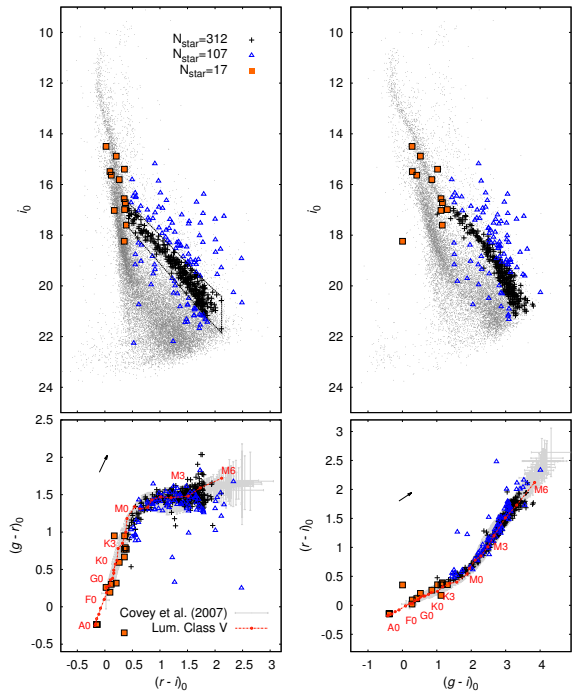}
  \caption{Color-magnitude and color-color diagrams for all detected flare stars.  Since our interest is particularly focused on the flare stars in the pre-defined cluster region (dashed line), the cluster\rq{s} extinction coefficient is used to deredden $gri$ magnitudes.  We adopt $A_{\lambda}/A_{V}$ values for $gri$ magnitudes that derived by \citet{gir04}.  The bluer ($(r-i)_{0} \le 0.43$) and redder ($(r-i)_{0} > 0.43$) stars outside the cluster region are marked with orange and blue dots, respectively.  The red lines represent the main-sequence stellar locus with solar metallicity \citep{cov07}.}
  \label{Fig5}
\end{figure}

\subsection{Rejection of false positives}
There is a possibility of false positives.  Since the new light curves are derived based on the total flux within a measuring aperture, various contaminations can cause photometric bias.  Most of false positives turned out to be related to CCD charge bleed (i.e., blooming), diffraction spikes around the bright stars, moving objects, and seeing-correlated variations due to blends of stars.  This kind of false positives can be most accurately verified by visual inspections.  Figure \ref{Fig3} shows thumbnail light curves and images of falsely identified events.  The abrupt variations in the photometry are evident when artifacts intrude into the measuring aperture of target stars. 

As shown in the top panels of Figure \ref{Fig4}, flare star V2172 is an interesting case; a real flare and a false-positive event (moving object) are observed at different epochs.  The latter one is easily discriminated from real flare with its symmetric light curve.  Another exceptional case is a pair of flare candidates (V2233 and V2234), which exhibit a sudden increases in brightness that would be expected in a typical flare.  However we note that one of them is caused by the flare of the adjacent star, not its intrinsic magnetic activity (see bottom panels of  Figure \ref{Fig4}).  Since these objects are very faint and their position in the master frame is very close ($0.79^{\prime\prime}$), it is difficult to distinguish in a single frame.  Further difference imaging analysis is needed to confirm which star is really responsible for this flare event.

\subsection{Summary of flare search}
After visual inspection of all candidates found in the region of interest, as defined in the Section 2.1, we identified a total of 604 unique flare events from M-type dwarfs.  Among these events, 420 flares are from 312 stars in the cluster sample, while 184 flares are from 107 stars in the field sample.  As shown in the two CMDs of Figure \ref{Fig5}, most of flare stars lie on the cluster sequence as expected (black pluses).  We show for comparison the flare sample of 17 F--K type dwarfs (21 flares) that were already reported in our previous paper \citep{cha15b} with orange squares in Figure \ref{Fig5}.  It shows that stars of all colors cooler than spectral type A can produce serendipitous flares, which are energetic enough to be observed even in the $r$-band.

\begin{deluxetable}{rrrrrrrrrrr}
\tabletypesize{\footnotesize}
\tablecolumns{12}
\tablewidth{0pt}
\tablecaption{Summary of flare samples\label{Tab2}}
\tablehead{
\colhead{}    &  \multicolumn{3}{c}{Cluster sample} &   \colhead{}   &
\multicolumn{3}{c}{Field sample} & \colhead{} & \colhead{} & \colhead{} \\
\cline{2-4} \cline{6-8}\\
\colhead{$r_{0}$\tablenotemark{a}} & \colhead{$N_\mathrm{s}$\tablenotemark{b}} & \colhead{$N_\mathrm{f}$\tablenotemark{c}} & \colhead{$N_\mathrm{e}$\tablenotemark{d}} & \colhead{} & \colhead{$N_\mathrm{s}$} & \colhead{$N_\mathrm{f}$} & \colhead{$N_\mathrm{e}$} & \colhead{} & \colhead{$\sum N_\mathrm{f}$} & \colhead{$\sum N_\mathrm{e}$}}
\startdata
16--17\phn & \nodata & \nodata & \nodata & & 72\phn & 5\phn & 10\phn & & 5\phn & 10\phn\\ 
17--18\phn & 148 (81)\phn & 17 (10)\phn & 25 (15)\phn & & 115\phn & 10\phn & 22\phn & & 27\phn & 47\phn\\
18--19\phn & 289 (175)\phn & 28 (18)\phn & 34 (23)\phn & & 258\phn & 16\phn & 33\phn & & 44\phn & 67\phn\\
19--20\phn & 295 (141)\phn & 32 (16)\phn & 53 (24)\phn & & 926\phn & 21\phn & 36\phn & & 53\phn & 89\phn\\
20--21\phn & 416 (142)\phn & 68 (25)\phn & 101 (36)\phn & & 1700\phn & 19\phn & 33\phn & & 87\phn & 134\phn\\
21--22\phn & 569 (80)\phn & 88 (13)\phn & 113 (14)\phn & & 2352\phn & 10\phn & 17\phn & & 98\phn & 130\phn\\
22--23\phn & 568 (1)\phn & 77 (0)\phn & 91 (0)\phn & & 3502\phn & 19\phn & 25\phn & & 96\phn & 117\phn\\
23--24\phn & 201 (0)\phn & 2 (0)\phn & 3 (0)\phn & & 3859\phn & 7\phn & 8\phn & & 9\phn & 11\phn\\
24--25\phn & 9 (0)\phn & \nodata & \nodata & & 529\phn & \nodata & \nodata & & \nodata & \nodata\\
\tableline
Total\phn & 2495 (620)\phn & 312 (82)\phn & 420 (112)\phn & & 13313\phn & 107\phn & 184\phn & & 419\phn & 604\phn\\
\enddata
\tablecomments{Extinction-corrected magnitudes are not reliable for the field region samples.  For the cluster sample, the values in the parenthesis correspond to the number of stars with membership information ($P_{mem} \geq 0.2$).}
\tablenotetext{a}{Average $r_{0}$ magnitude after extinction correction (see Section 2.1).}
\tablenotetext{b}{Number of sample stars in each magnitude bin.}
\tablenotetext{c}{Number of identified flaring stars, some of which flare multiple times during the observation span.}
\tablenotetext{d}{Number of flare events.}
\end{deluxetable}
  
Table \ref{Tab2} summarizes the number of flare stars ($N_{f}$), the number of flare events ($N_{e}$), and the number of sample stars ($N_{s}$) for nine consecutive magnitude intervals.  For the cluster sample, the values in the parenthesis correspond to the number of stars with membership information ($P_{mem} \geq 0.2$).  The number of flaring stars is only a small fraction of the total number of objects in the M37 field ($<3\%$), but it provides sufficient numbers for a statistical comparison of the characteristics.  For the cluster sample, the ratio of flare stars to sample stars ($N_{f}/N_{s}$) in each magnitude bin is about 10\%--16\%.  This ratio is similar to that observed in stars with $P_{mem} \geq 0.2$ (6\%--17\%).  The number of flare stars and their occurrence rate both increase with decreasing luminosity down to magnitude limit of the survey ($r\sim23$).  These are likely due to the increasing contrast effect of the white-light flare emission against the cool photosphere (e.g., \citealt{wal11,dav12}).  Since the detectability of flare events depends on several observational factors, we discuss limitations of our data quality that affect the range of marginally detectable flare events (see Section 5.2 for details).  

Because flare stars have finite activity lifetimes and their activity fraction declines with Galactic height (e.g., \citealt{wes08,wes11}), our field sample is expected to be a part of nearby, young, active K--M dwarfs in a thin disk.  Assuming a extinction coefficient given by \citet{sch98}, we estimated the distance of each star derived from the photometric parallax relations (see \citealt{boc10} for details): 
\begin{equation}
m_{r} - M_{r}(r - i) = 5\log d - 5 + A_{r},
\end{equation} where $d$ is the distance, $m_{r}$ is the apparent magnitude, $A_{r}$ is the extinction in the $r$-band, and $M_{r}(r - i)$ is the color-absolute magnitude relation\footnote{$M_{r}$ = $5.025 + 4.548(r-i)+0.4175(r-i)^{2}-0.18315(r-i)^{3}$; (see erratum in \citealt{boc10})}.  After obtaining the distance, the vertical distance from the Galactic plane, $\vert Z \vert$, can be estimated by converting a spherical coordinate system ($l, b, d$) to a cylindrical coordinate system ($R, Z, \phi$).  The position of the Sun was set at $R_\sun$=8.5 kpc and $Z_\sun$=15 pc above the mid-plane of the Galactic disk.  The results show that most of them are likely to lie within 200 pc from the Galactic plane.  In spite of the intrinsic large uncertainty of photometric parallaxes, this supports that our field sample stars are indeed foreground, thin disk population.

\begin{figure*}[!t]
\centering
   \subfloat[]{\includegraphics[width=0.85\linewidth, angle=0]{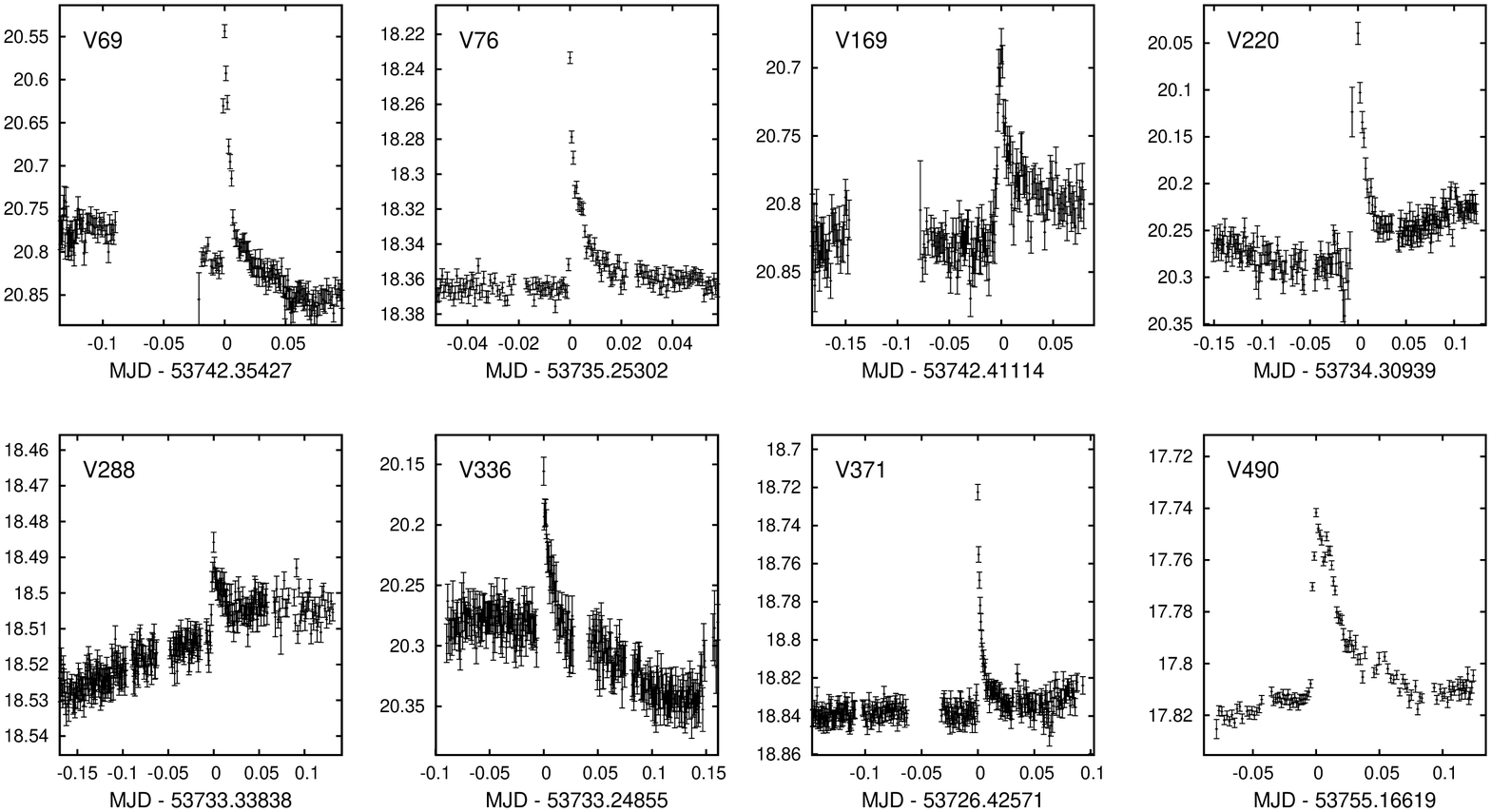}}\vspace*{-1.3em}
   \subfloat[]{\includegraphics[width=0.85\linewidth, angle=0]{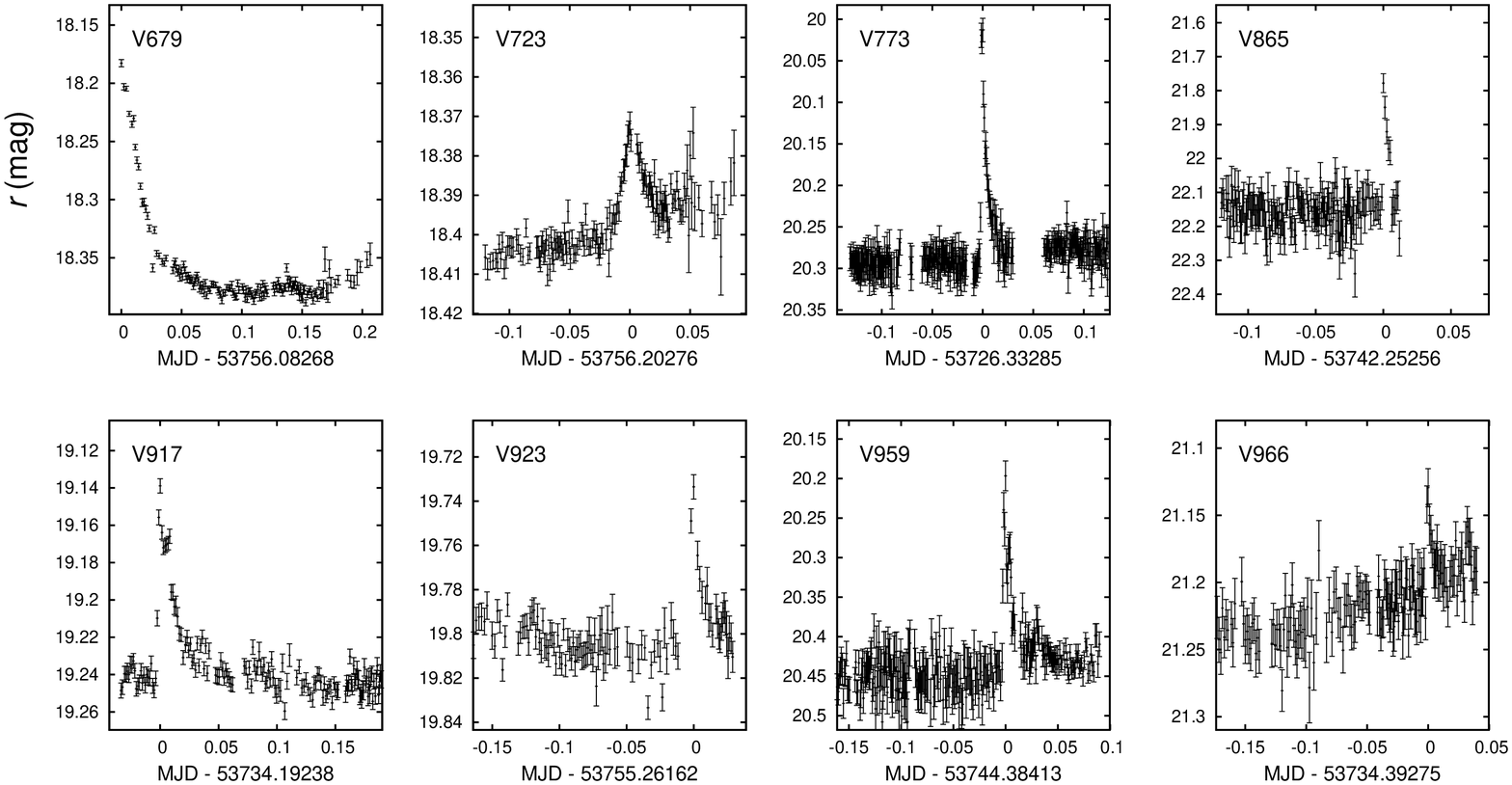}}\vspace*{-1.3em}
   \subfloat[]{\includegraphics[width=0.85\linewidth, angle=0]{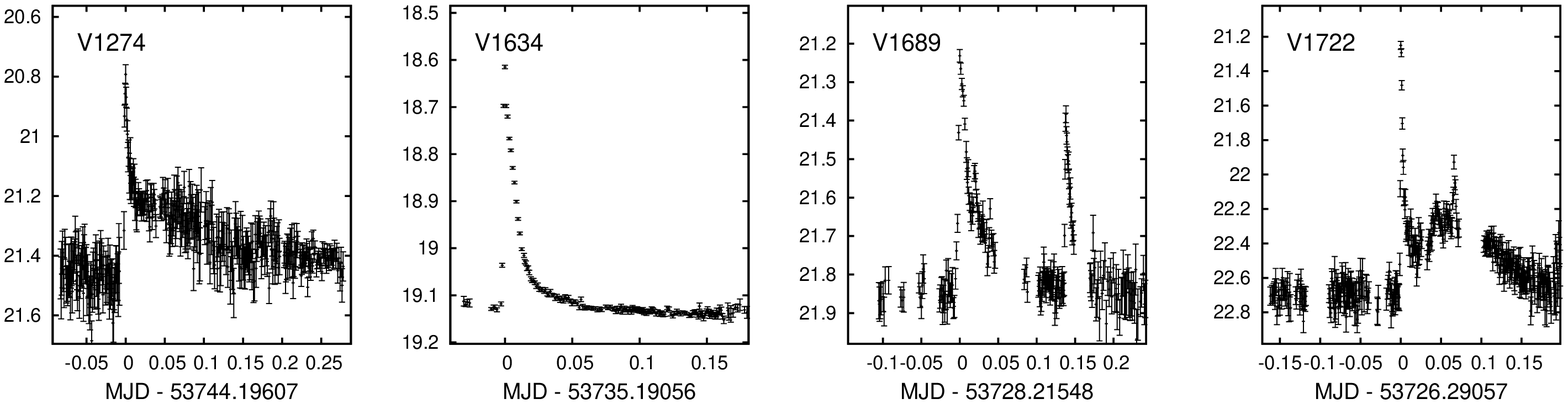}}\vspace*{-1.3em}
   \caption{Examples of flare light curves before the removal of underlying periodic variations.  One of the fascinating aspects of flares is their diverse morphological shapes.}
  \label{Fig6}
\end{figure*}  

\section{Measurement of flare parameters in the light curves}
As shown in Figure \ref{Fig6}, the light curves of discovered flare events show diverse morphologies in terms of its shape, size, and duration.  Flare variability cannot be described as a one-parameter family of model functions, so we obtain several observational and physical parameters from flare light curves in Section 4.1 and Section 4.2.

\subsection{Observational flare parameters}
We define basic observational parameters of flare events as follows: (i) the magnitude of observed peak ($\Delta m_{peak}$) and its epoch ($t_{peak}$), (ii) the timescales of flare evolution both in the pre-peak ($t_{q,rise}$) and the post-peak ($t_{q,decay}$) regimes of flare curve (i.e., the parts before and after the point of maximum), where the subscript $q$ denotes the flux fraction relative to peak level, and (iii) the magnitude difference $\Delta m_{base}$ between the first epoch data ($t_{0,rise}$) and the last epoch data ($t_{last}$) of flare observations.  For the convenience of calculation, we take the epoch of maximum flare light as a reference point.  

\begin{figure}[!t]
\begin{center}
  \includegraphics[width=\linewidth, angle=0]{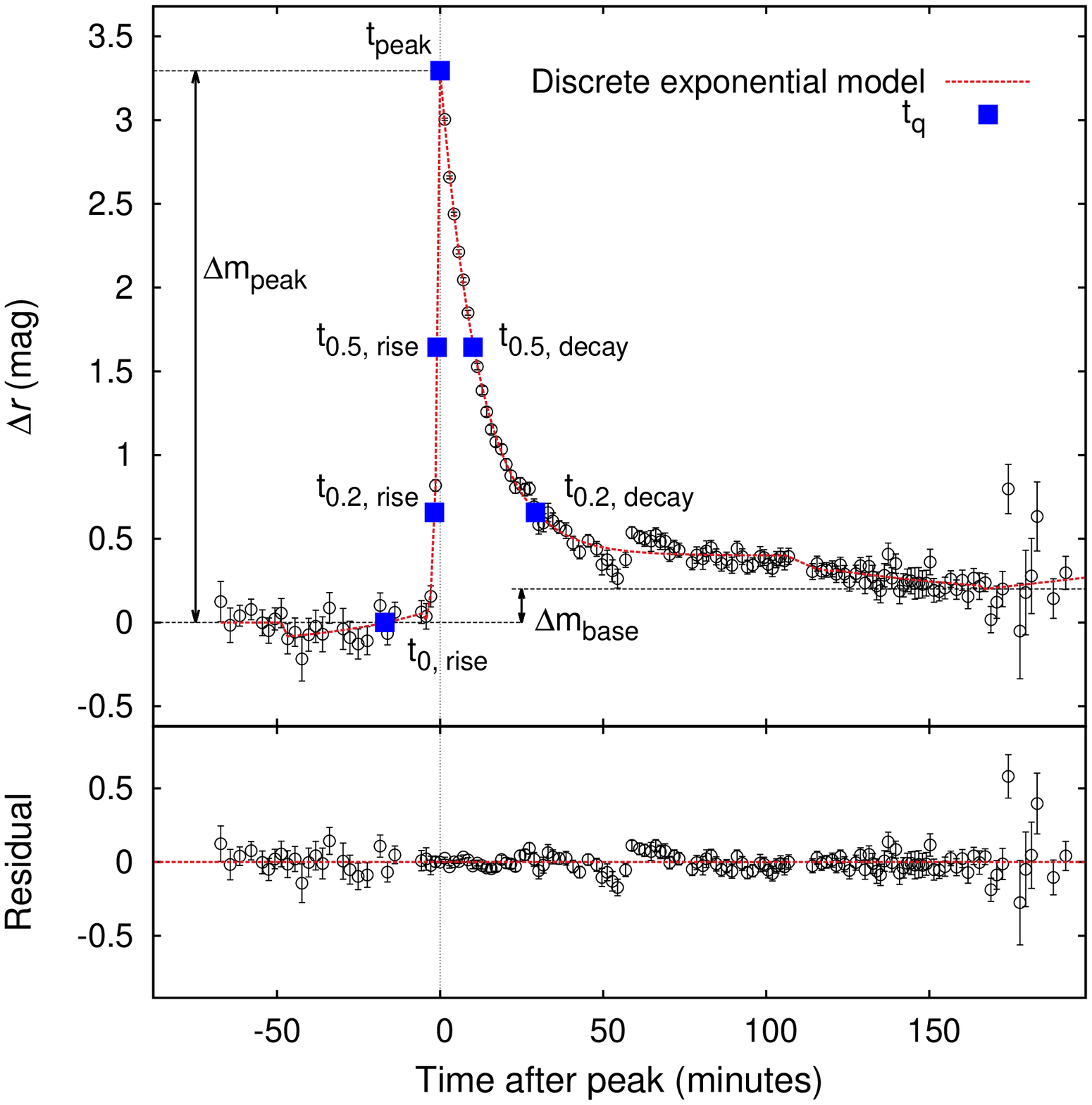}  
  \caption{Schematics of observational flare parameters in the light curves.  For the convenience of calculation, the zero-point of the $x$-axis is shifted to the peak time ($t_{peak}$) of light curve, while in the $y$-axis we set the zero-point to lie at $\Delta$$r = 0$ after subtracting the mean magnitude of light curve.  By using the discrete exponential model (red line), we defined a time-related parameter, $t_{q}$, which can be either positive (e.g., $t_{0.5, decay}$, $t_{0.2, decay}$) or negative (e.g., $t_{0.5, rise}$, $t_{0.2, rise}$).  In this plot, we also derived the flare amplitude ($\Delta m_{peak}$) and the magnitude difference ($\Delta m_{base}$) between the first and the last data point for a given subset.}
  \label{Fig7}
\end{center}
\end{figure}

The concept of observational flare parameters is illustrated for a typical case in Figure \ref{Fig7}.  The moment of flare peak is considered as a reference point ($t$ = 0).  We initially determined the timescales of flare variability in the range of $0 \le q \le 1$, using single or multiple discrete exponential decay models.  The model is expressed as the sum of the exponential functions:  
\begin{equation}
y(t) = \sum_{k} f(t; \alpha_{k},\beta_{k},\lambda_{k}),  
\end{equation}
\begin{displaymath}
f(t; \alpha_{k},\beta_{k},\lambda_{k}) = \alpha_{k} \exp(-\lambda_{k} \cdot t) + \beta_{k},
\end{displaymath} where $\alpha_{k}$, $\beta_{k}$, and $\lambda_{k}$ are the best-fitting parameters to produce piecewise curve approximation.  The fitting is performed with an iterative optimization method based on the Levenberg-Marquardt solver\footnote{\url{http://www.gnu.org/software/gsl/}}.  In this way, we grow each curve piece by adding data points until the fitting error for that piece exceeds a certain threshold (i.e., $\chi_\nu^{2} \gg 1$).  In the time interval $t_{0, rise} \le t \le t_{0,decay}$, we measure the timescale parameter $t_{q}$ by analyzing the original flare curve with a linear interpolation method that does not assume a functional form. 

The remaining parameters are obtained from the processed light curves.  We list the observational parameters of the whole flare samples in Table \ref{TabA1}.  Using these parameters, we can easily reproduce template flare light curves of a wide range of morphologies.  In some cases, the timescales of total flare duration can not be specified by a single parameter $\tau_{0}$ $(= t_{0,rise} + t_{0,decay}$) because observations did not cover the entire phase of flare evolution.  This is why we introduce additional $\tau_{q}$ parameters (e.g., $\tau_{0.5}$, $\tau_{0.2}$, $\tau_{0.1}$) to characterize timescales of flares (see Section 5.1).

\subsection{Physical flare parameters}
The physical parameters of each flare were only obtained for the cluster sample.  Following \citet{ger72} and \citet{mof74}, we first calculated the equivalent duration $P_{r}$.  This quantity can be thought of as the time interval over which the quiescent star emits as much energy as was released during the duration of the flare \citep{wal11}.
\begin{equation}
P_{r} = \int{\left(\frac{I_{0+f}(t)}{I_{0}} - 1 \right) dt},
\end{equation} where $I_{0}$ is the flux of the star in its quiescent state and $I_{0+f}$ is the flaring flux.  In order to simplify the procedure, we regard groups of flares or sub-peaks within a flare as one flare event.  The flare energy $E_{r}$ is expressed as the product of the quiescent luminosity of the star $L_{r}$ in $r$-band and the equivalent duration of the flare:

\begin{equation}
E_{r} = L_{r} \times P_{r}.
\end{equation}

To estimate the quiescent luminosity for each star in the cluster sample, we adopt the distance of 1490 pc \citep{har08}.  Since the measured flux density within the passband corresponds to the stellar flux at the effective wavelength of the filter, it can be approximated as:
\begin{eqnarray}
L_{r} & = & 4 \pi d^{2} \int_{r} f_{\lambda} d\lambda \nonumber \\
				& \simeq & 4 \pi d^{2} \cdot f_{\lambda_\mathrm{eff}} \cdot \Delta \lambda,
\end{eqnarray} where $d$ is the distant of the star in pc, $f_{\lambda}$ is the spectral density of flux in erg cm$^{-2}$ s$^{-1}$ $\mathrm{\AA^{-1}}$, $\lambda_\mathrm{eff}$ is the effective wavelength of the $r$-filter, and its bandwidth $\Delta \lambda$.  The filter system used for the {\sc MMT/Megacam} is subtly different from those of the {\sc SDSS} \citep{fuk96}.  The response function of {\sc Megacam} $r$-filter is more extended to the red part of spectrum than that of the {\sc SDSS}, and thus the integrated spectral flux is somewhat larger (about 0.3\%) for our data set.  Errors in the estimated quiescent luminosities are dominated by the distance uncertainty ($\pm$ \unit[120]{pc}) which causes variation of about 15\% (i.e., $\log L_r = \pm0.07$).  For fainter stars $\log L_{r}$ $<$ 29.93 erg s$^\mathrm{-1}$, the photometry error becomes dominant. 

We also estimated the flare luminosity $L_{r, peak}$ at maximum brightness (erg s$^\mathrm{-1}$) after subtracting the quiescent stellar $r$-band flux.  The derived  physical parameters of flares detected in the cluster sample are listed in Table \ref{TabA2}.  Since the observations do not always have a continuous time coverage, only a lower limit to the physical parameters could be given in some cases ($\sim$20\% of our full sample).  Hence, subsequent estimates should be interpreted with some caution when we statistically quantify the properties of flares, both including or excluding these subsamples with incomplete light curve.

\section{Statistical properties of flare timescale, energy, and frequency}

\subsection{Light curve characteristics of flares}
In this section, we discuss the temporal and peak characteristics of flares in the whole flare star sample (Table \ref{TabA1}).

\begin{figure}[!t]
  \includegraphics[width=0.94\linewidth, angle=0]{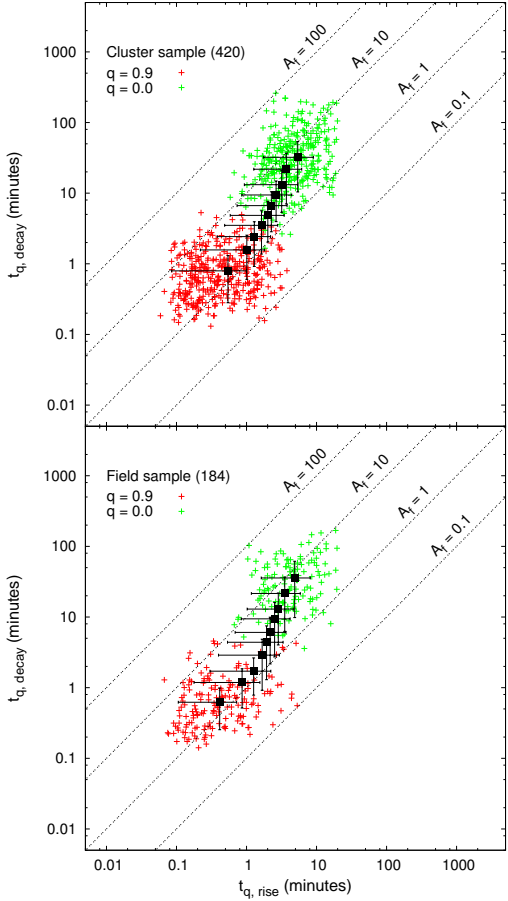}
  \caption{Time evolution of the individual flares on the $t_\mathrm{q, rise}$--$t_\mathrm{q, decay}$ diagram.  The black squares show the clipped mean and standard deviation of for $q$ values from 0.0 to 0.9 in steps of 0.1.  For convenience, the distributions of timescales are shown at both $q=0$ (green points) and $q=0.9$ (red points), respectively.  The dashed lines indicate flare asymmetry $A_{f}$ between 0.1 and 100.}
  \label{Fig8}
\end{figure}

\begin{figure}[!t]
  \includegraphics[width=\linewidth, angle=0]{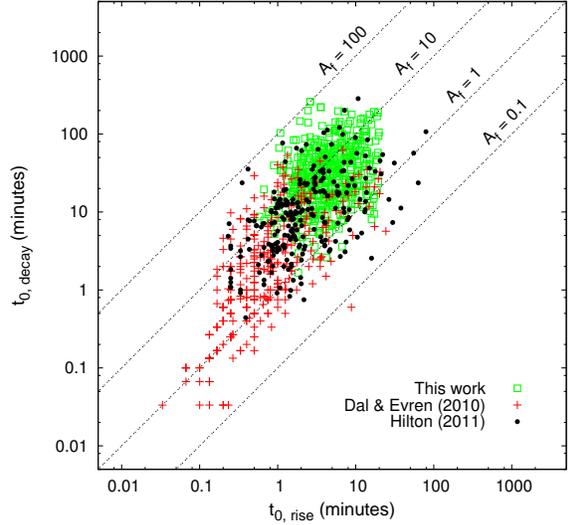}
  \caption{Comparison of the rise and decay timescales between this work (green square) and the two recent works of flare monitoring.  More than two hundreds of flares were detected by either photoelectric observations \citep{dal10} or by photometric observations \citep{hil11}, respectively.  Overall, there is a good agreement with the distributions of flare timescales obtained from previous studies.  Significant short-term (less than $\tau_{0}$ = 1 minute) flares were only observed by photoelectric observations.}
  \label{Fig9}
\end{figure}

\subsubsection{Rise and decay timescales}
The time evolution of stellar flares is characterized by different timescales, but it is often convenient to treat them as two principal phases: an initial impulsive phase and later gradual phase.  This reflects the basic assumption that most stellar flares follow similar consequence of the same elementary physical mechanisms \citep{ben10}.  The impulsive phase is defined as the quickly varying part of the light curve which shows a fast rise to maximum magnitude followed by a fast decay.  The gradual decay phase begins with a turnover from fast to slow decay, as suggested by \citet{haw91,kow13}.  In general, the initial phase of energy release lasts from tens of seconds to tens of minutes, and then returns to its pre-event level on timescales of tens of minutes or hours. 

In Figure \ref{Fig8}, we show the time evolution of individual flares on the flare rise $t_{q,rise}$ and decay $t_{q,decay}$ timescales.  The rise timescales $t_{0,rise}$ of flares are distributed between 0.6 and 20 minutes, while the decay timescales are longer than the rise timescales ($t_{0,decay}$ = 1.6--260.1 minutes); and the total duration $\tau_{0}$ ranges from 3.5--263 minutes (0.06--4.4 h).  For each $q$ value, we used the rise and decay timescales to measure the flare asymmetry, which is defined as the fraction of timescales after and before the flare peak ($A_{f}$ = $t_{q,decay}/t_{q,rise}$).  We take the values between 0.1 and 100, in which $A_{f} = 1$ corresponds to a time-symmetric case ($t_{q,decay} = t_{q,rise}$).  It is shown that the boundary between impulsive and gradual phases occurs at $q$ = 0.5--0.6, where it is shown for the values of $A_{f}$ = 2--3.  In practice, \citet{kow13} used the full width of the light curve at half-maximum (i.e., $\tau_{0.5}$) as the timescale of the impulsive phase of the flare.  In their flare samples, the impulsive timescale covers a very broad range with a mean value of $<$$\tau_{0.5}$$>$ = $7.1\pm6.3$ minutes.

In Figure \ref{Fig9}, we compare our rise and decay timescales with two recent work of flare monitoring \citep{dal10,hil11}.  Their observations provide a large, statistically significant sample of flares of a few active stars.  Considering the coarse time resolution of our data, there is a general agreement with flare timescales obtained from previous studies.  These characteristic timescales are common for both cluster and field flare stars, indicating that they follow the same physical processes during flares.  Our sample tend to have slightly longer rise and decay timescales compared to individual flares from a few stars.  Since the flare duration is tightly correlated with the flare energy, our sample tend to be more energetic (see also Figure \ref{Fig16} and Figure \ref{Fig17}).  \citet{dal10} found many more short time-scale flares with their much finer time resolution.  This result tells us the approximate minimum time-scale for which the present study is sensitive.

\subsubsection{Peak amplitudes}
The photometric response of flares depends on spectral type of the star and filters.  Using a two-component flare model with quiescent M0--M6 spectral templates, \citet{dav12} predicted the red-optical and NIR response of flares in M dwarfs, and showed that red-optical filters ($gri$ bands) are sensitive to medium and large flares.  Following the method described in \citet{dav12}, we estimated the predicted changes in $u$-band for our $r$-band flares\footnote{J. R. A. Davenport provides a simple IDL program to compute the $ugrizJHK_{s}$ amplitudes of a flare given the spectral type of the star and the desired amplitude in any one filter.  A release version of the code is now available at \url{ https://github.com/jradavenport/flare-grid}.}.  By assigning photometric spectral types using colors with the covariance matrix technique presented in \citet{kow09}, we can transform the $\Delta r$-band to the $\Delta u$-band amplitudes for all flares in the cluster sample.  

\begin{figure}[!t]
\centering
  \includegraphics[width=\linewidth, angle=0]{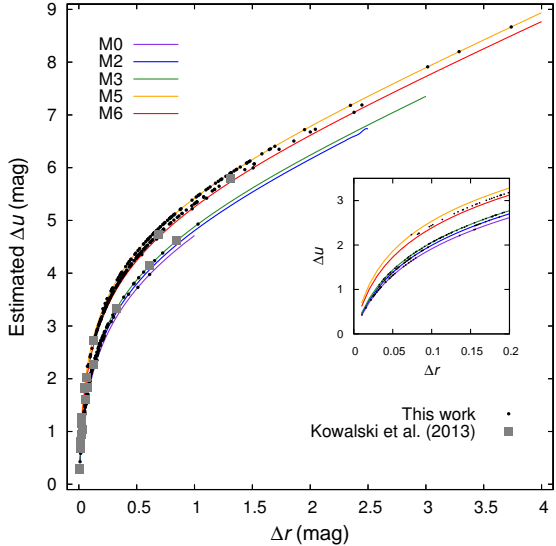}
  \caption{Transformation between the $r$-band response and the $u$-band response for each flare in the cluster sample.  The gray squares indicate the sample of \citet{kow13}.  We estimated the $r$-band amplitudes of flares for their sample by applying the same conversion method.  The subfigure on the lower-right shows the zoom-in of the low-amplitude flares ($\Delta r < 0.2$ mag).}
  \label{Fig10}
\end{figure} 

\begin{figure*}[!t]
\centering
   \subfloat[]{\includegraphics[width=0.85\linewidth, angle=0]{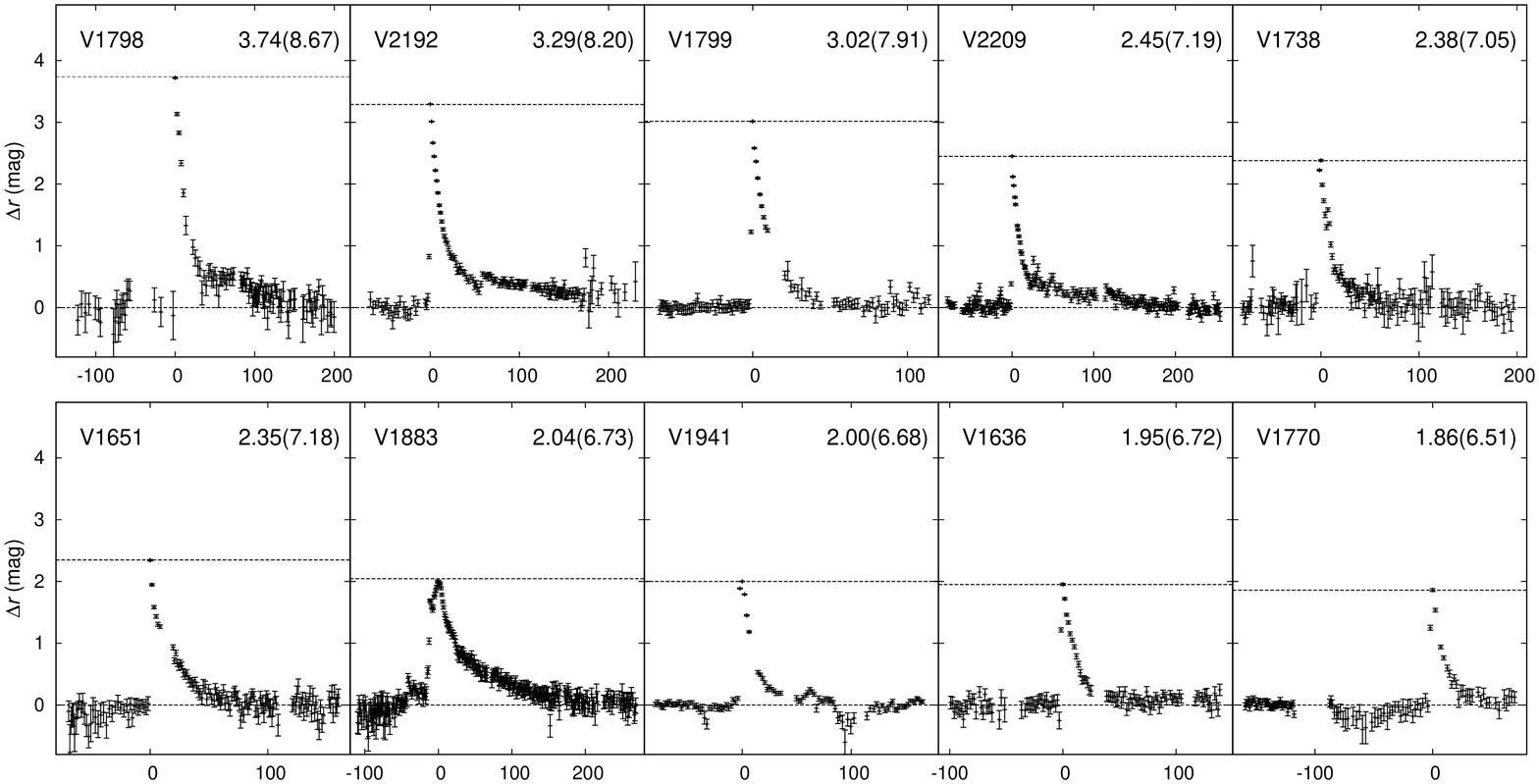}}\vspace*{-1.7em}
   \subfloat[]{\includegraphics[width=0.85\linewidth, angle=0]{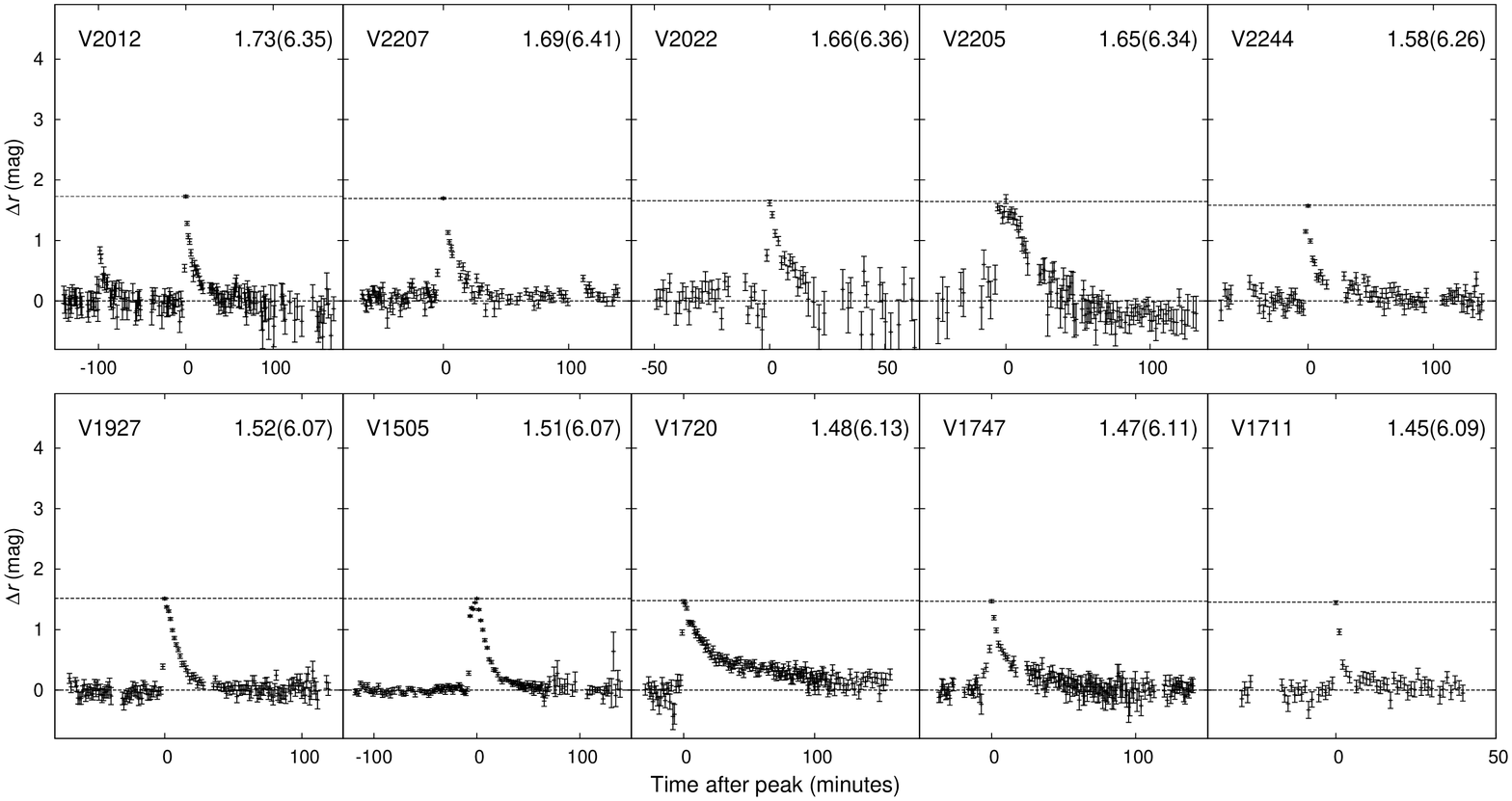}}\vspace*{-1.5em}
   \caption{Light curves for the flares with inferred $\Delta u > 6$ mag in the cluster sample, which are arranged in the order of decreasing peak amplitude (dashed lines).  The $r$-band and its inferred $u$-band amplitudes are also given in the plot.}
  \label{Fig11}
\end{figure*}

\begin{deluxetable}{ccccccrrccc}
\tabletypesize{\tiny}
\tablecolumns{11}
\tablewidth{0pt}
\tablecaption{Properties of the flare samples with inferred $\Delta u >$ 6 mag\label{Tab3}}
\tablehead{
\colhead{} & \colhead{} & \colhead{} & \colhead{} & \colhead{$r$} & \colhead{$\Delta r$ ($\Delta u$)} & \colhead{$\tau_{0}$} & \colhead{log $E_{r}$} & \colhead{Period\tablenotemark{b}} & \colhead{$P_{mem}$\tablenotemark{c}} & \colhead{}\\
\colhead{VarID} & \colhead{StarID} & \colhead{FlareID} & \colhead{Sp Type\tablenotemark{a}} & \colhead{(mag)} & \colhead{(mag)} & \colhead{(minute)} & \colhead{(erg)} & \colhead{(days)} & \colhead{(\%)} & \colhead{XID\tablenotemark{d}}
}
\startdata
V1798\phn & 055205.47+324035.05\phn & F1\phn & M5\phn & 23.51\phn & 3.74 (8.67)\phn & 142.75\phn & 34.018\phn & \nodata\phn & \nodata\phn & \nodata\phn \\
V2192\phn & 055256.65+322457.53\phn & F1\phn & M5\phn & 22.96\phn & 3.29 (8.20)\phn & $>180.43$\phn & $>34.143$\phn & \nodata\phn & \nodata\phn & \nodata\phn \\
V1799\phn & 055205.56+323821.31\phn & F1\phn & M5\phn & 22.85\phn & 3.02 (7.91)\phn & 84.61\phn & 33.862\phn & \nodata\phn & 0.00\phn & 168\phn \\
V2209\phn & 055301.57+322821.00\phn & F1\phn & M4\phn & 22.62\phn & 2.45 (7.19)\phn & 201.49\phn & 33.968\phn & 0.32\phn & 0.08\phn & \nodata\phn \\
V1738\phn & 055158.37+322327.37\phn & F1\phn & M6\phn & 23.32\phn & 2.38 (7.05)\phn & 78.16\phn & 33.653\phn & \nodata\phn & \nodata\phn & \nodata\phn \\
V1651\phn & 055144.84+322901.99\phn & F1\phn & M5\phn & 23.30\phn & 2.35 (7.18)\phn & 69.35\phn & 33.618\phn & \nodata\phn & \nodata\phn & \nodata\phn \\
V1883\phn & 055215.13+323535.84\phn & F2\phn & M4\phn & 21.89\phn & 2.05 (6.73)\phn & $>151.25$\phn & $>34.482$\phn & \nodata\phn & 0.00\phn & 318\phn \\
V1941\phn & 055222.57+323317.73\phn & F1\phn & M4\phn & 21.56\phn & 2.00 (6.68)\phn & 94.32\phn & 34.209\phn & \nodata\phn & 0.59\phn & 473\phn \\
V1636\phn & 055142.63+322928.22\phn & F1\phn & M5\phn & 22.53\phn & 1.95 (6.72)\phn & 74.22\phn & 33.727\phn & \nodata\phn & \nodata\phn & \nodata\phn \\
V1770\phn & 055202.47+322831.46\phn & F1\phn & M4\phn & 22.43\phn & 1.86 (6.51)\phn & 58.33\phn & 33.701\phn & \nodata\phn & 0.05\phn & \nodata\phn \\
V2012\phn & 055231.97+323747.11\phn & F2\phn & M4\phn & 22.28\phn & 1.73 (6.35)\phn & 53.43\phn & 33.557\phn & \nodata\phn & \nodata\phn & \nodata\phn \\
V2207\phn & 055301.49+322821.62\phn & F1\phn & M5\phn & 23.65\phn & 1.69 (6.41)\phn & 53.20\phn & 33.142\phn & \nodata\phn & \nodata\phn & \nodata\phn \\
V2022\phn & 055232.83+322906.18\phn & F1\phn & M5\phn & 22.96\phn & 1.66 (6.36)\phn & 21.56\phn & 33.202\phn & \nodata\phn & \nodata\phn & \nodata\phn \\
V2205\phn & 055301.28+324038.13\phn & F1\phn & M5\phn & 23.54\phn & 1.65 (6.34)\phn & 38.64\phn & 33.401\phn & \nodata\phn & \nodata\phn & \nodata\phn \\
V2244\phn & 055309.20+323800.11\phn & F1\phn & M5\phn & 22.94\phn & 1.58 (6.26)\phn & 86.58\phn & 33.286\phn & \nodata\phn & \nodata\phn & \nodata\phn \\
V1927\phn & 055221.06+323912.63\phn & F1\phn & M4\phn & 21.96\phn & 1.52 (6.07)\phn & 64.12\phn & 33.722\phn & 0.94\phn & 0.22\phn & \nodata\phn \\
V1505\phn & 055121.71+323647.04\phn & F2\phn & M4\phn & 22.39\phn & 1.51 (6.07)\phn & 64.95\phn & 33.771\phn & \nodata\phn & 0.002\phn & \nodata\phn \\
V1720\phn & 055156.68+322701.14\phn & F2\phn & M5\phn & 22.87\phn & 1.48 (6.13)\phn & 139.62\phn & 33.778\phn & \nodata\phn & \nodata\phn & \nodata\phn \\
V1747\phn & 055159.86+324407.19\phn & F1\phn & M5\phn & 23.24\phn & 1.47 (6.11)\phn & 82.97\phn & 33.277\phn & \nodata\phn & \nodata\phn & \nodata\phn \\
V1711\phn & 055155.09+322723.09\phn & F1\phn & M5\phn & 23.28\phn & 1.45 (6.09)\phn & 41.83\phn & 32.596\phn & \nodata\phn & \nodata\phn & \nodata\phn \\
\enddata
\tablenotetext{a}{Photometric spectral types assigned based on the Table 1 in \citet{kow13}.}
\tablenotetext{b}{Periods taken from \citet{cha15b}.}
\tablenotetext{c}{Membership probability based on CMD position and radial 
distance from the cluster center \citep{nun15}.}
\tablenotetext{d}{X-ray counterpart ID number from \citet{nun15}.}
\end{deluxetable}

As shown in Figure \ref{Fig10}, even small flares down to $\Delta u \sim 0.4$ are completely recovered by our survey.  We also note that our survey detects very large flares ($\Delta u > 6$ mag).  The flares with inferred $\Delta u > 6$ mag are listed in Table \ref{Tab3} and light curves are shown in Figure \ref{Fig11}, arranged in order of decreasing peak amplitude. Such large-amplitude flares have been rarely discovered around the mid- and late-M dwarfs (e.g., \citealt{haw91, kow09, kow10, sch14}).  They do exist in abundance.  It tells us that the $u$-band observation of the flares have the advantage of not requiring high-precision photometry, compared to our $r$-band observations.

\subsection{Statistical correlations of flare energy and other key parameters}
In this section, we investigate relations between flare energy and other derived parameters for flare stars in the cluster sample (Table \ref{TabA2}).  With an assumption that cluster stars of similar physical quantities (mass, luminosity, and age) should have similar statistical properties of flare activity, we group the flare stars into seven magnitude bins ranging  from $r_{0}=17$ to $r_{0}=24$.  This follows the ergodic hypothesis suggested by \citet{ger05}, for which the number statistics of flares on $n$ stars of similar brightness over the time $T$ is the same as those of the same star over the time $n T$.  Each group can also be considered as spectral types along the cluster main-sequence, in which decreasing magnitude indicates later spectral type.  In a larger sample of stars of similar spectral types, this statistical approach is useful to understand their flaring nature (e.g., \citealt{kow09, hil11, wal11, ost12, dav12, shi13}).

\begin{figure*}[!t]
\centering
   \subfloat[]{\includegraphics[width=0.5\linewidth, angle=0]{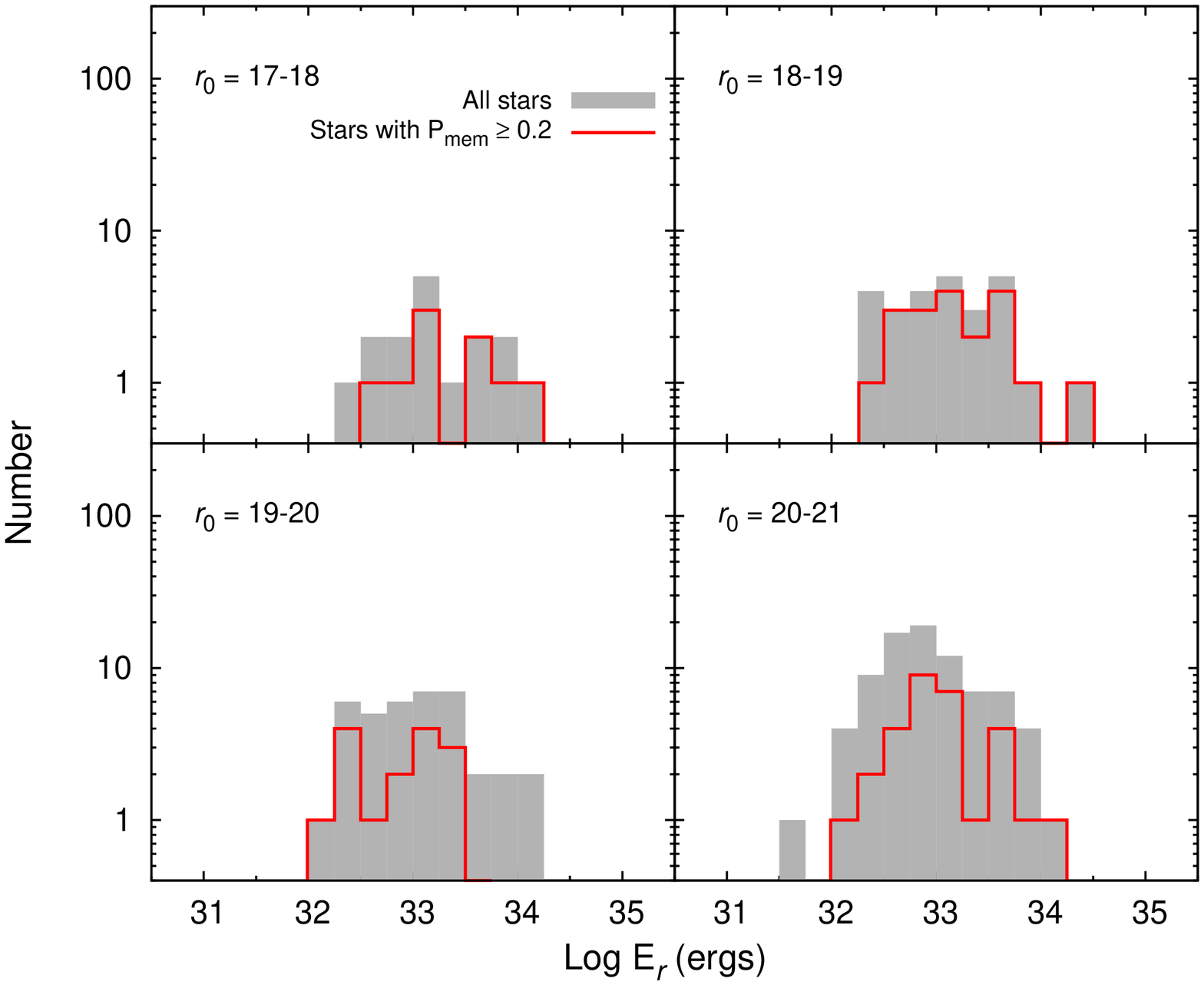}}
   \subfloat[]{\includegraphics[width=0.5\linewidth, angle=0]{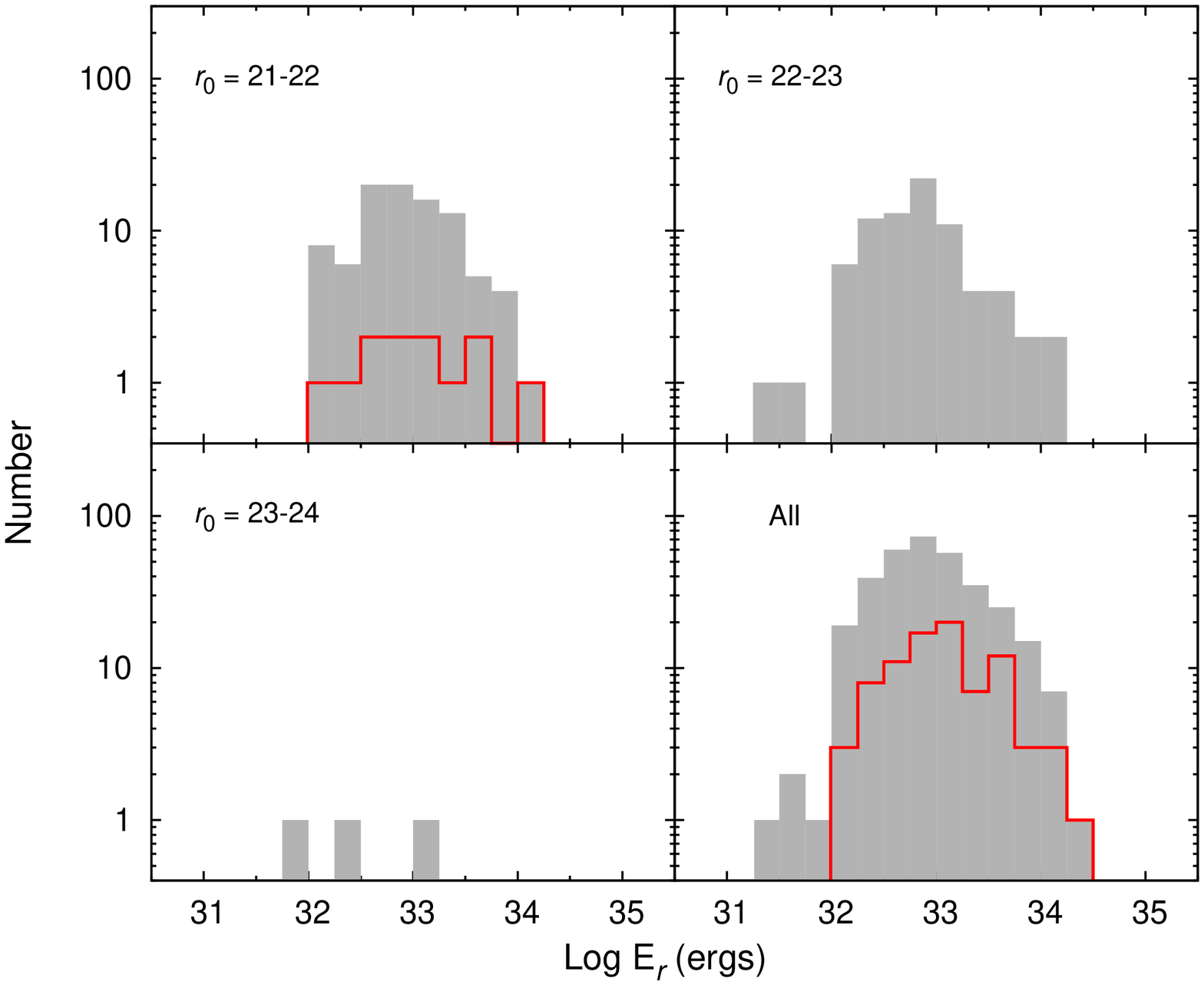}}\vspace*{-1.3em}
   \caption{Number distribution of the total flare energy in $r$-band (log $E_{r}$) per magnitude interval (gray histograms).  The red histograms represent stars with membership information ($P_{mem} \geq 0.2$).  The last panel shows the distribution of the total energy for all flares in the cluster sample with complete information ($\sim$80\% of our full sample).}
   \label{Fig12}
\end{figure*}

\subsubsection{Distribution of flare energy $E_{r}$}
It is known that stellar flares radiate energy at all wavelengths like on the Sun.  Among these wavelengths, continuum emission emitted by white-light flare is the main contributor to the total radiated energy \citep{haw91}.  The energy budget based on the $r$-band light curve is therefore not sufficient to approximate the total flare energy.  In Figure \ref{Fig12}, we show the distribution of $r$-band flare energy ($E_{r}$) for the samples with complete information, which ranges from 2.9$\times10^{31}$ to 2.4$\times10^{34}$ erg.  We also overlay the energy distributions of cluster stars with $P_{mem} \geq 0.2$ (red histogram).  These two distributions are indeed similar, but it is difficult to compare them directly with $r_{0} > 21$ due to the lack of cluster membership information.  We find that the upper limit on the flare energy becomes at least 7 times larger ($>$1.69$\times$$10^{35}$ erg) if we include the subsamples with incomplete light curve.  The histograms seem to show a turnover value at $E_{r}\sim$ 7.6$\times$$10^{32}$ erg, indicating that the fall-off in the low energy side is likely due to incompleteness of our sample.  It appears to be difficult to directly use these distributions for testing accurate occurrence rates of flares with a typical power-law shape.
   
We found many flares with energies above $10^{33}$ erg, called superflares, which is about 10--100 times more energetic than the largest known solar flare.  Such flares are rarely reported in previous ground-based observations, but it is now known that there is a large number of late-type (G--M) dwarf stars that show superflares using Kepler data \citep{mae12,shi13,can14}.  Their results suggest that superflares occur more frequently on young and/or K--M type stars, which is a consequence of the age-activity-rotation relation.  Thus, superflares with energies above $10^{34}$ erg might well be possible for much younger cluster stars than the M37.

For comparison purpose, we converted all flare energies to the commonly used passband (Johnson $U$ filter).  There is a simple conversion relation among flare energies measured in different filters \citep{lac76,kil78}.
\begin{equation}
E_{U} = 1.2 E_{B} = 1.8 E_{V} = 2.1 E_{8050},
\end{equation} where $E_{U, B, V}$ are the flare energies in the Johnson $UBV$ passbands and $E_{8050}$ is a flare energy in the passband with an effective wavelength of 8050 $\AA$ (close to the Cousins $I_{c}$-band).  However, our filter does not overlap precisely with wavelengths covered by either $V$-band or $I_{c}$-band.  To obtain the relation between $E_{U}$ and $E_{R}$, we used results of the flare energy budget from a multi-wavelength observing campaign on the flare star AD Leo \citep{haw91,haw03}.  For the available sample of 4 flares with well-measured light curves, we derived the conversion relations.  These relations are nearly matched to previous measurements in the literatures except for the $V$-band energy budget.  We found that $E_{U} \simeq 1.05 \pm 0.41 E_{R}$.  However, we caution that flare energy scaling between $E_U$ and red passbands may vary due to the presence and strength of a red continuum component in the used flare sample \citep{kow13}.  Further conversion to MMT/Megacam $r$ band is determined by the ratio of the FWHM of the two filters, $E_{R} \sim 1.38 E_{r}$, following the approach of \citet{ost12}.  This relation gives an approximate relation between $E_{U}$ and $E_{r}$:
\begin{equation}
E_{U} \simeq 1.45 E_{r}
\end{equation} where we assume that flares with a wide range of total emitted energy follow the same relation.  The effects of 1-$\sigma$ variation in the scaling factor between $E_{U}$ and $E_{R}$ lead to a range of about $\pm0.1$ in $\log E_U$.
 
\begin{figure}[t]
  \centering
  \includegraphics[width=\linewidth, angle=0]{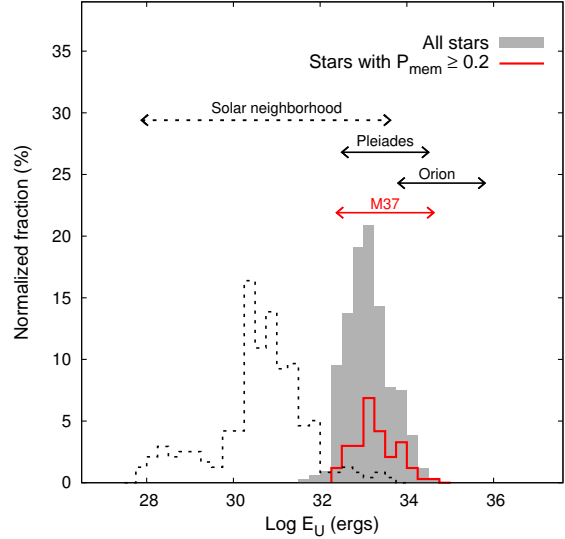}  
  \caption{Observed ranges of estimated $U$-band flare energy for this work and for similar studies from the literature.  The gray and red histograms are all cluster sample and samples with membership information ($P_{mem} \geq 0.2$), respectively.  The dashed histogram is for the field M dwarfs by \citet{hil11}.  For flare stars in the Orion and Preiades clusters, the observed energy ranges are taken from \citet{ger83}.  Flares on M dwarfs in open clusters are more energetic than field M dwarfs in the Solar neighborhood within 25 pc.}
  \label{Fig13}
\end{figure}

We plotted the observed ranges of estimated $U$-band flare energy for three open clusters and Solar neighborhood (Figure \ref{Fig13}).  The flare energy range that can be observed in each study is limited by detection sensitivity.  Although M37 is the most distant cluster among the three clusters, our study includes relatively weak flares.


\subsubsection{Relation between flare energy and peak amplitude}
During the phase of maximum brightness, the flare peak spectra exhibit a steeply rising continuum toward NUV wavelength regime (Balmer continuum: BaC) and its white-light continuum.  The shape of this emission resembles that of a hot blackbody emission with a temperature near 10,000 K, regardless of its morphological type, peak luminosity, or total energy (see Section 6 in \citealt{kow13} for details).  It is believed to be originated from a compact region on the stellar surface, such as the footprint of magnetic loops on the Sun.  Since the Stefan-Boltzmann law for blackbody radiation depends only on the blackbody temperature, the amplitude (or energy rate) of observed flux continuum should be closely correlated with fractional area coverage for flares \citep{haw03,kow10}.  Accordingly, the ratio of the projected area of the flaring region to the visible stellar surface is expected to be maximized during the flare peak.  The characteristic size of this area is called the filling factor of the blackbody component, $X_\mathrm{BB}$, that can be derived from the spectra or often from the white-light continuum using broadband photometry (e.g.,\citealt{haw92,kow13}).  

\begin{figure}[!t]
   \subfloat[]{\includegraphics[width=\linewidth, angle=0]{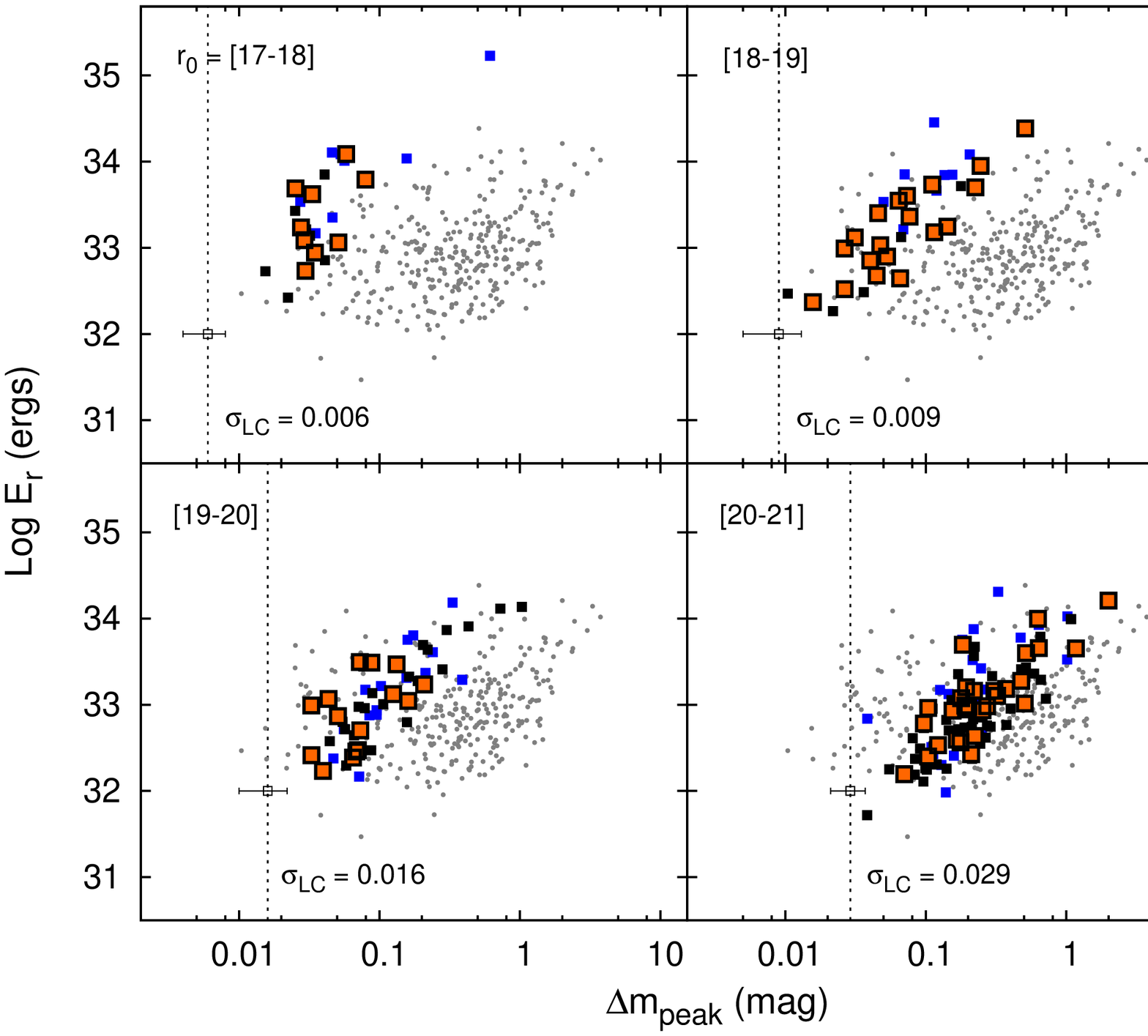}}\vspace*{-1.3em} 
   \subfloat[]{\includegraphics[width=\linewidth, angle=0]{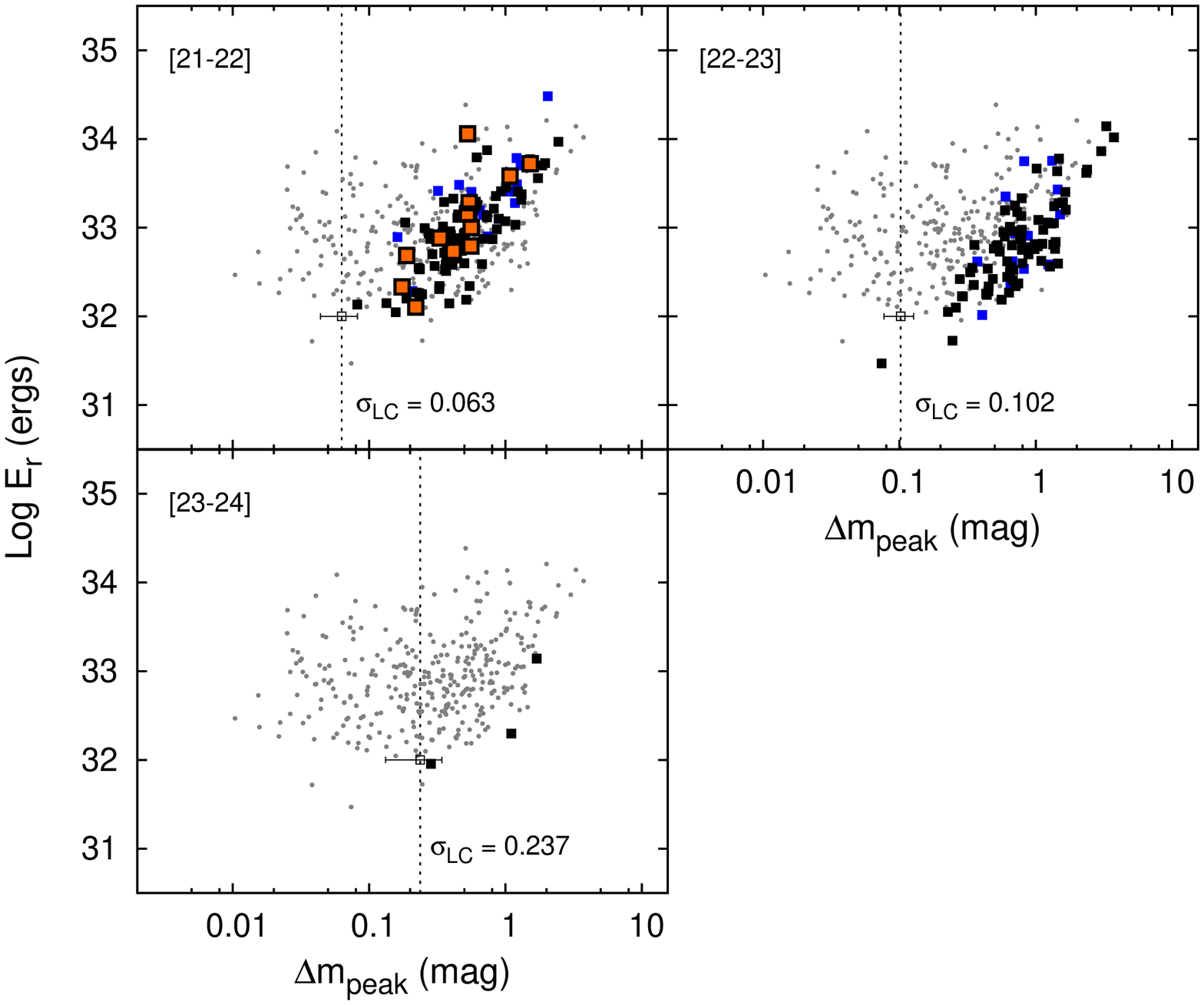}}\vspace*{-1.3em}
   \caption{Scatter plot of flare amplitude ($\Delta m_\mathrm{peak}$) vs. flare energy (E$_{r}$) for each magnitude bin.  The gray points and black squares represent the full sample of detected flare events and subsample in each magnitude bin, respectively.  We also show the lower limit for the subsamples with incomplete light curve (blue squares).  For comparison, the orange squares show stars with membership information ($P_{mem} \geq 0.2$).  The vertical dashed lines with open square indicate the mean RMS of all light curves ($\sigma_\mathrm{LC}$) in that magnitude bin.}
   \label{Fig14}
\end{figure}

Figure \ref{Fig14} shows a scatter plot of flare amplitude against its energy of our sample for each magnitude bin.   The flare energy is related to the peak amplitude in log scale, $E_{r} \propto \Delta m_\mathrm{peak}$.  For a given energy value, the later-type stars exhibit relatively larger amplitude variations.  The origin of this relation is simply due to the contrast effect as mentioned in previous section.  It is interesting that the distribution of subsamples with incomplete light curve (blue squares) does not bias the result.  Meanwhile, the distributions of flare amplitude provide bounds on the flare energy that should be detected by our flare-search method.  The minimum detectable energy of our survey can be determined by several parameters: apparent magnitude of the star, photometric measurement uncertainty, or light curve RMS value in each segment.  Among them, the mean RMS of all light curves ($\sigma_\mathrm{LC}$) in each magnitude bin is the major limiting factor in the effort of detecting weak flares with less than $10^{32}$ erg.

\begin{figure}[!t]
  \centering
  \includegraphics[width=\linewidth, angle=0]{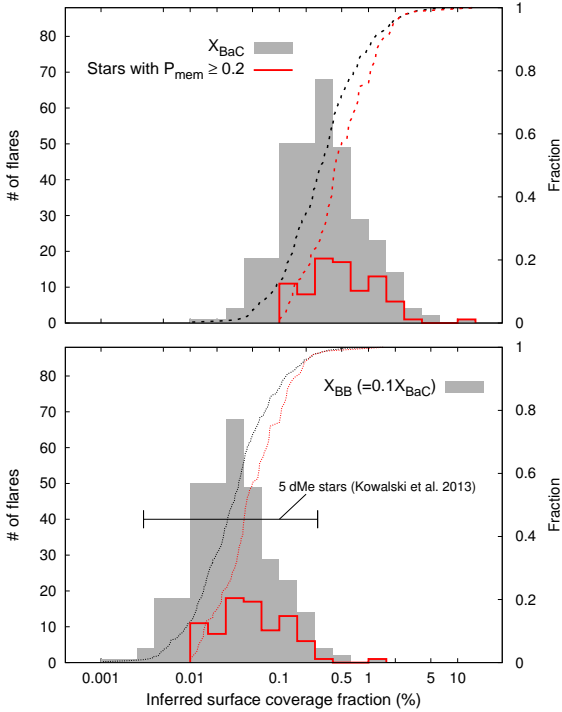}  
  \caption{Estimated surface coverage of flares for the Balmer continuum (top) and the blackbody component (bottom) using the two-component flare model.  The gray and red histograms are all cluster sample and samples with membership information ($P_{mem} \geq 0.2$), respectively. The dashed lines show the cumulative distributions.  The fractional area coverage for the blackbody varies between approximately 0.002--0.5\% during the flare peak.  The observed range of the filling factors derived from the flare spectra of five active M3--M4 dwarfs \citep{kow13}  is indicated by a horizontal bar in the bottom panel.}
  \label{Fig15}
\end{figure}

Since we have no spectral information for the flares from our survey, we used the two-component flare model of \citet{dav12} to estimate the maximum areal coverages of each flare peak.  The flare model assumes that the overall shape of flare spectral energy distribution follows a 10,000 K blackbody continuum, and the size of Balmer continuum is set to have 10 times larger surface area coverage than the blackbody component ($X_\mathrm{BB}$=0.1$X_\mathrm{BaC}$).  Using the flaregrid data, we chose to do the reverse (spline) interpolation between the surface coverage fraction ($X_\mathrm{BaC}$) and the simulated flux enhancement ($\Delta r$) in the photometric band.  The results of estimated flaring area for the two emission components are shown in Figure \ref{Fig15}, in which we also show for comparison (see the bottom panel) the $X_\mathrm{BB}$ values obtained from the flare peak spectra \citep{kow13}.  The filling factors of the blackbody emission are 0.002--0.5\% of visible stellar hemisphere at flare peak.  The physical sizes of the flaring area are given by the product of the corresponding surface area (i.e., $\pi R^{2}$) of the star.  These values are calculated assuming a effective radius of 0.62, 0.49, 0.44, 0.39, 0.26\footnote{The original reference provided by \citet{rei05} states 0.36, but should be 0.26.}, 0.20, and 0.15 $R/R_{\sun}$ for M0--M6 dwarfs, respectively.  The inferred results show that the size of flare area ($2\times10^{17}$--$2.5\times10^{21}$ cm$^{2}$) is considerably larger than the value ($\sim5.8\times10^{17}$ cm$^{2}$; see Table A1 in \citealt{nei83}) for solar white-light flares at flare maximum.  These characteristics are consistent with previous observations of stellar flares (e.g., \citealt{van96,haw03,kow10,kow13}).

\begin{figure}[!t]
   \subfloat[]{\includegraphics[width=\linewidth, angle=0]{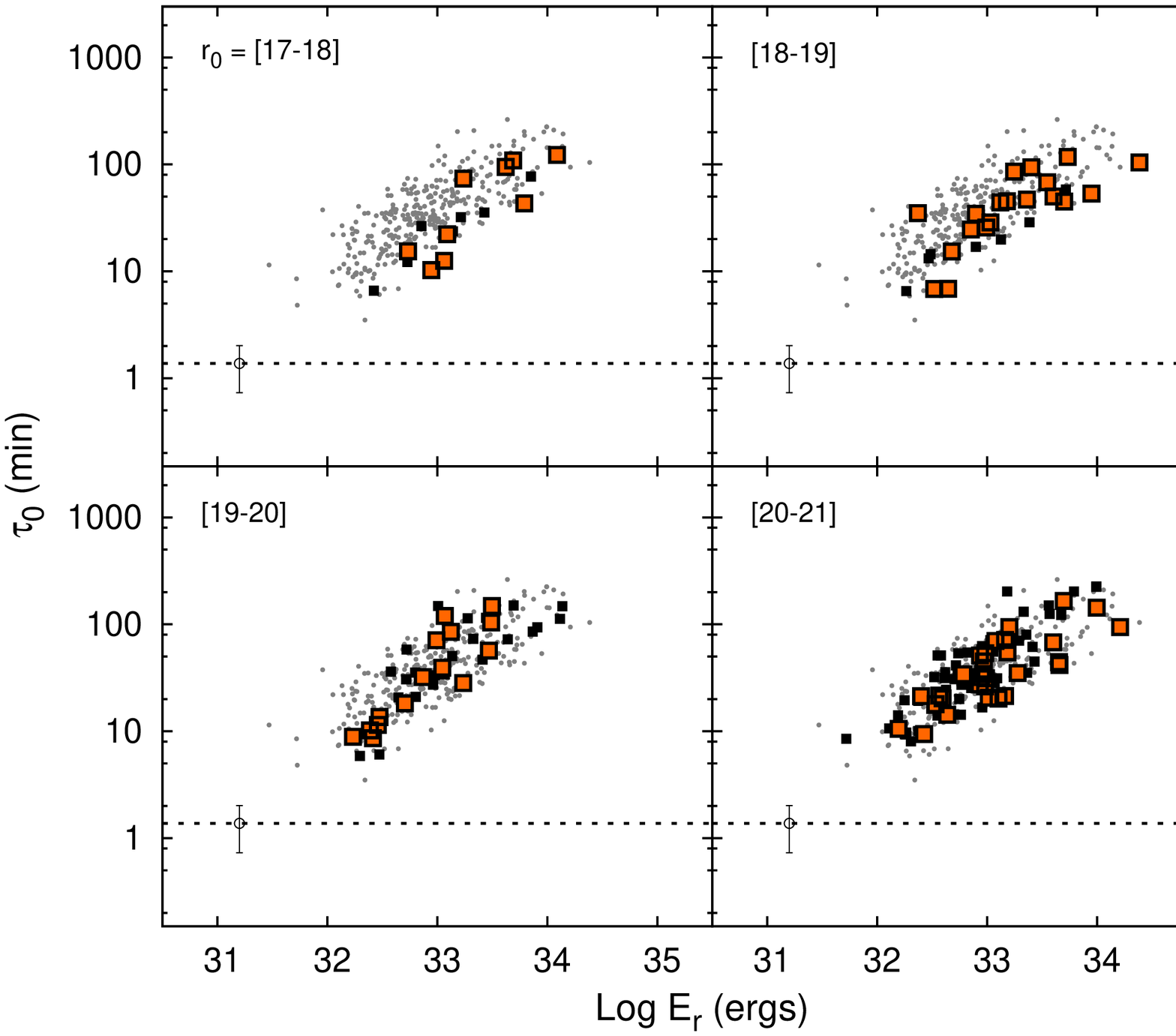}}\vspace*{-1.3em}
   \subfloat[]{\includegraphics[width=\linewidth, angle=0]{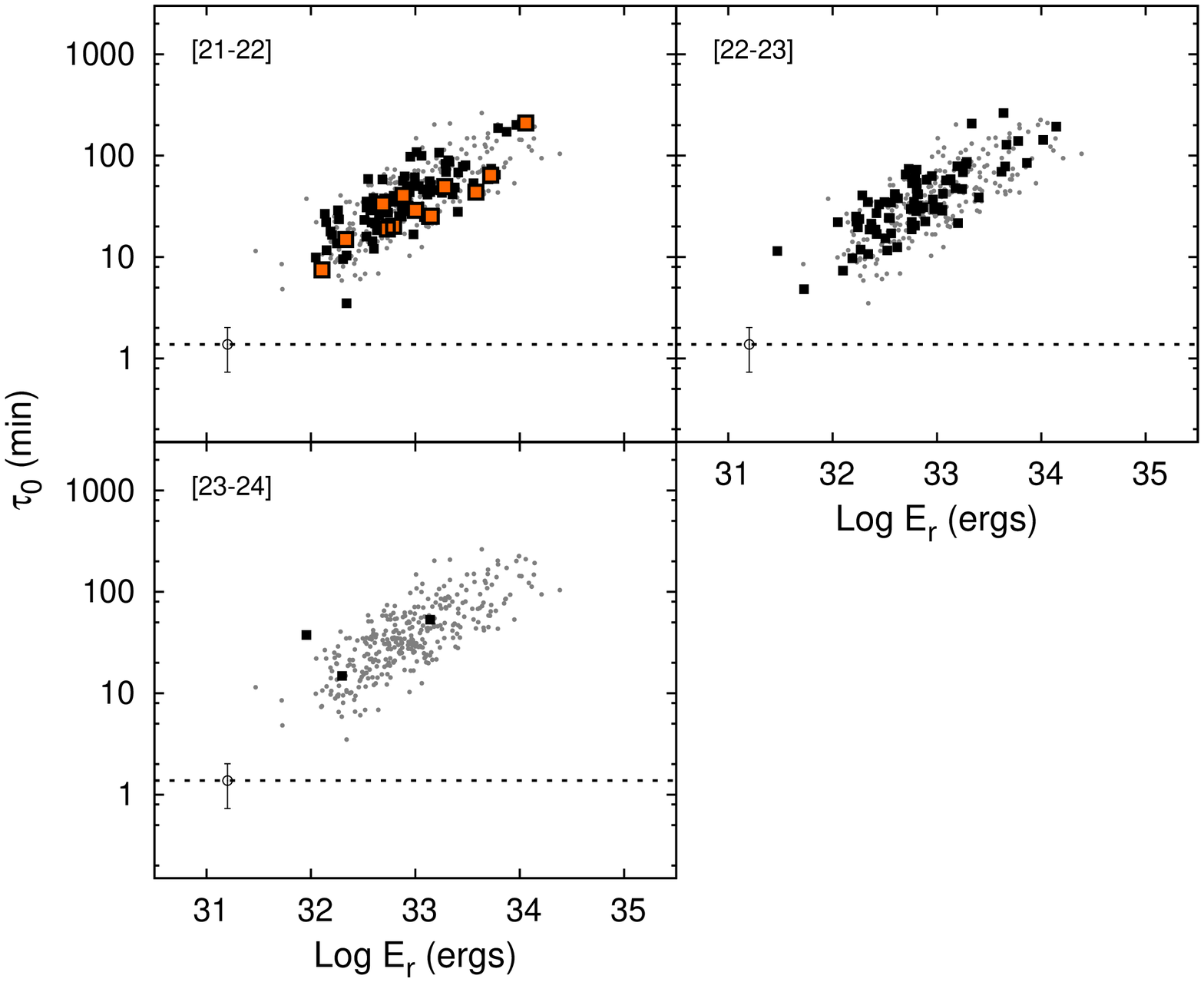}}\vspace*{-1.3em}
   \caption{Scatter plot of flare energy ($E_{r}$) vs. its duration ($\tau_\mathrm{0}$) for each magnitude bin.  The gray points and black squares represent the full sample of detected flare events and subsample in each magnitude bin, respectively.  For comparison, the orange squares show stars with membership information ($P_{mem} \geq 0.2$).  The dashed horizontal line indicates the average time interval between two consecutive exposures ($\sim$$82\pm38$ seconds).}
  \label{Fig16}
\end{figure}

\subsubsection{Relation between flare energy and temporal parameter} 
We investigate the correlation between flare energy and its temporal parameters, such as duration, rise time, and decay time.  The $E_{r}$--$\tau_{0}$ relation is limited by the time resolution of our survey.  In measuring a flare and its duration, we need at least 2 points ($<$ 3 minutes) to detect a flare by our detection method, but we find that data sets with more measurements are required to obtain its total duration (at least 5 data points; 7 minutes).  The cut-off at short time-scales and low energies shown in Figure \ref{Fig16} is likely due to our time resolution.
        
\begin{figure}[!t]
  \includegraphics[width=\linewidth, angle=0]{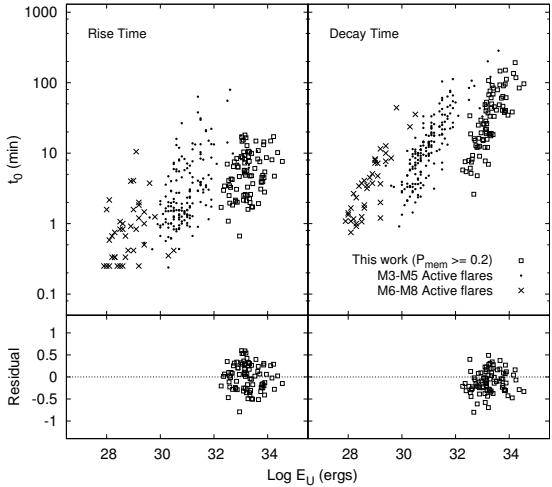}  
  \caption{Rise and decay times for the flares from the active, mid-M dwarfs (M3--M5) and active, late-M dwarfs (M6--M8), and this work.  The former two samples are adopted from \citet{hil10}.  The $E_{r}$--$\tau_{0}$ relation seen in the Figure \ref{Fig16} is mostly explained by the decay time of flares.  Depending on the total energy release by flares, these three groups occupy the different areas on the energy-time plane.  There is little overlap between the flares from our work and the M3 through M5 active flares.} 
  \label{Fig17}
\end{figure}

Figure \ref{Fig16} also shows a tight correlation in log-log diagrams of the flare energy and its duration in all magnitude intervals. This trend is very similar for the $E_{r}$--$t_{0,decay}$ relation (see right panel of Figure \ref{Fig17}).  In order to obtain the relation in a form of $y=ax+b$, we perform a least-squares regression analysis for log $E_{r}$ versus log $\tau_{0}$ and log $t_{0, decay}$ in unit of seconds.
\begin{equation}
\log \tau_{0} = (0.57\pm0.05) \log E_{r} - (15.61\pm1.57),
\end{equation}
\begin{equation}
\log t_{0, d} = (0.66\pm0.06) \log E_{r} - (18.53\pm1.86),
\end{equation} which shows good agreement within the uncertainties.  It is an inevitable result of flare time evolution since a large fraction of the flare energy is emitted during the gradual phase.  Thus, the range in flare energy can be estimated by these equations with a good level of significance.  On the other hand, the correlation is less tight for the rise times than the decay times as reported by the previous studies (e.g., \citealt{pet89,hil11}).
\begin{equation}
\log t_{0, r} = (0.21\pm0.07) \log E_{r} - (4.47\pm2.19).
\end{equation}  These general relations are similar, but not identical, to those of the $U$-band observations (e.g., \citealt{let97}).  

\subsection{Flare occurrence rate}
\begin{figure}[!t]
  \includegraphics[width=1.0\linewidth, angle=0]{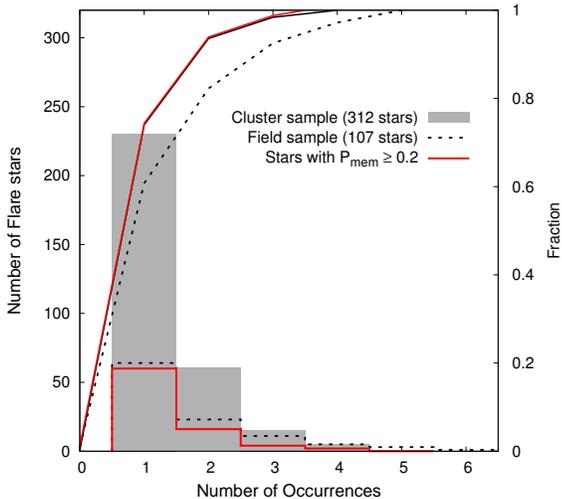}
  \caption{Number of flare stars as a function of flare occurrences.  Most of flares appear only once in the cluster (gray histogram) and field (black histogram) samples, respectively.  For comparison, the red histogram shows stars with membership information ($P_{mem} \geq 0.2$).}
  \label{Fig18}
\end{figure}

According to the intensive photometric monitoring of a single active star (e.g., \citealt{mof74, ish91}), flares occur rather randomly in time as a Poisson process, and high-energy flares are less frequent.  Unfortunately, we have very few samples ($\leq 4$ recurrences) from each flare star to address statistical properties of individual star (Figure \ref{Fig18}).  Using the same scheme mentioned in Section 5.2, however, we estimate the frequency with which a flare of a particular energy can occur, and then compare these rates with previous works.

\begin{deluxetable}{ccccccccccc}
\tabletypesize{\scriptsize}
\tablecaption{Mean flare number rate ($\mathcal{N}_{R}$) observed in the cluster sample\label{Tab4}}
\tablewidth{0pt}
\tablehead{
\colhead{} & \colhead{} & \colhead{} & \colhead{} & \colhead{} & \colhead{} & \colhead{} & \colhead{} & \colhead{} & \colhead{$\mathcal{N}_{R}(N_{f})$} & \colhead{$\mathcal{N}_{R}(N_{s})$} \\
\colhead{$r_{0}$} & \colhead{$N_\mathrm{s}$} & \colhead{$N_\mathrm{f}$} & \colhead{$N_\mathrm{e}$} & \colhead{$n_{1}$} & \colhead{$n_{2}$} & \colhead{$n_{3}$} & \colhead{$n_{4}$} & \colhead{$n_{5}$} & \colhead{($h^{-1}$)} & \colhead{($h^{-1}$)}
}
\startdata
17--18\phn &   148 (81)\phn &  17 (10)\phn & 25 (15)\phn &  11 (7)\phn & 4 (1)\phn & 2 (2)\phn & \nodata & \nodata & 0.020 (0.020)\phn & 0.0023 (0.0025)\phn \\
18--19\phn &   289 (175)\phn &  28 (18)\phn & 34 (23)\phn &  22 (13)\phn & 6 (5)\phn & \nodata & \nodata & \nodata & 0.016 (0.017)\phn & 0.0016 (0.0018)\phn \\
19--20\phn &   295 (141)\phn &  32 (16)\phn & 53 (24)\phn &  18 (10)\phn & 9 (5)\phn & 3 (0)\phn & 2 (1)\phn & \nodata & 0.022 (0.020))\phn & 0.0024 (0.0023)\phn \\
20--21\phn &  416 (142)\phn &  68 (25)\phn & 101 (36)\phn & 45 (18)\phn & 16 (4)\phn & 5 (2)\phn & 1 (1)\phn & 1 (0)\phn & 0.020 (0.019)\phn & 0.0033 (0.0034)\phn \\
21--22\phn &   569 (80)\phn &  88 (13)\phn & 113 (14)\phn & 69 (12)\phn & 15 (1)\phn & 2 (0)\phn & 2 (0)\phn & \nodata & 0.017 (0.015)\phn & 0.0027 (0.0024)\phn \\
22--23\phn & 568 (1)\phn & 77 (0)\phn & 91 (0)\phn & 65 (0)\phn & 10 (0)\phn & 2 (0)\phn & \nodata & \nodata &  0.016 (\nodata)\phn & 0.0021 (\nodata)\phn \\
23--24\phn &   201 (0)\phn & 2 (0)\phn & 3 (0)\phn & 1 (0)\phn & 1 (0)\phn & \nodata & \nodata & \nodata &  0.020 (\nodata)\phn & 0.0002 (\nodata)\phn \\
\tableline 
Total  & 2495 (620)\phn & 312 (82)\phn & 420 (112)\phn & 231 (60)\phn & 61 (16)\phn & 14 (4)\phn & 5 (2)\phn & 1 (0)\phn & 0.018 (0.018)\phn & 0.0023 (0.0024)\phn \\
\enddata
\tablecomments{The values in the parentheses are for stars with membership information ($P_{mem} \geq 0.2$).}
\end{deluxetable}

\subsubsection{Flare number rate}
We first estimate the average flare number rate for each magnitude bin:
\begin{mathletters}
\begin{equation}
\mathcal{N}_{R} (N_\mathrm{f}) = \frac{N_\mathrm{e}}{\tau_\mathrm{obs} \times N_\mathrm{f}},
\end{equation} 
\begin{equation}
\mathcal{N}_{R} (N_\mathrm{s}) = \frac{N_\mathrm{e}}{\tau_\mathrm{obs} \times N_\mathrm{s}},
\end{equation} 
where $N_\mathrm{e}$ is the number of flares, $N_\mathrm{s}$ is the number of observed stars, $N_\mathrm{f}$ is the number of stars that flare, and $\tau_\mathrm{obs}$ is the total monitoring time (=74.095 h).  The derived values are summarized in Table \ref{Tab4}.  The mean number rates of flares are 0.018 hr$^{-1}$ and 0.002 hr$^{-1}$ for flaring stars and all stars (i.e., flaring+non-flaring stars) in the cluster sample, respectively.  It is not much different that of samples with $P_{mem} \geq 0.2$ (see values in parentheses in Table \ref{Tab4}).  This kind of statistics has not been possible nor reported previously.  The rates are comparable or larger than those values (0.005 and 0.001 h$^{-1}$) found for field M2-–M9 dwarfs \citep{roc06}, but are much smaller than those of the active M dwarfs (e.g., \citealt{lac76,ish91}).  Comparison among different work can be misleading because the flare rate does not take into account the level of flare energy.
\end{mathletters}

Based on the observed flare stars and flare number rate in Table \ref{Tab4}, we can estimate the total number of flare stars in the cluster.  We follow the statistical analysis as described in \citet{amb75}.  The number of stars on which $k$ flares occurred over the time $t$ are derived under the following two assumptions: (i) the statistical occurrence frequency distributions of flares follow a Poisson distribution, and (ii) the mean rates of flare occurrence are the same for all stars in the cluster.  This is defined as following equation (Case I):
\begin{equation}
n_\mathrm{k} = N e^{-\nu t} \frac{(\nu t)^{k}}{k!},
\end{equation} where $N (= \sum_{k=0}^{\infty} {n_{k}}$) is the estimated total number of flare stars in the cluster and $\nu$ is the mean flare rate of them.  Using this simple estimation, the number of flare stars $n_{0}$ on which no flares occurred over the time $t$ can be calculated directly without any assumptions.
\begin{equation}
n_{0} = \frac{n_{1}^{2}}{2 n_{2}}.
\end{equation} However, there is no reason to suppose that the mean rate of flares should be the same for all flare stars.  When mean flare rates are different in each magnitude bin, the equation is replaced by the sum of Poisson distributions (Case II):
\begin{equation}
n_\mathrm{k} = \sum_{i} N_{i} e^{-\nu_{i} t} \frac{(\nu_{i} t)^{k}}{k!},
\end{equation} where $i$ is the number of groups with different mean rates of flares.  In this case, the value of $n_{0}$ can be expressed as below.
\begin{equation}
\frac{n_{1}^{2}}{2 n_{2}} \leq n_{0} \leq \frac{n_{1}^{2}}{n_{2}}.
\end{equation}

\begin{figure}[!t]
  \includegraphics[width=\linewidth, angle=0]{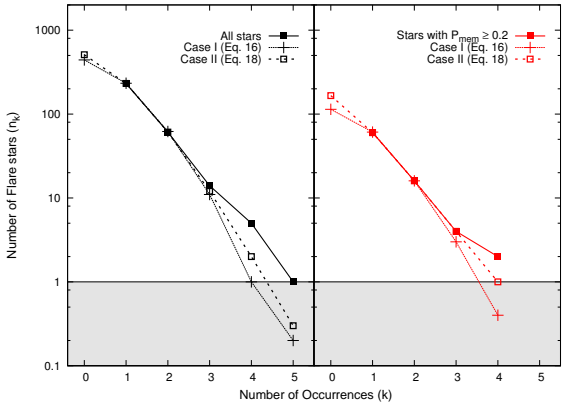}
  \caption{Same as Figure \ref{Fig18}, but for the model of case I (Equation 16) and case II (Equation 18), respectively.  The only difference between the two models is the assumption of occurrence rate of flares.  The gray shaded region indicates that for the non-detectable flares ($n_{k} < 1$).}
  \label{Fig19}
\end{figure}

For these two cases, we calculate the number of flare stars $n_{k}$ in which $k$ flares may have been observed, and then estimate the total number of flare stars $N$.  We obtained the two parameters $n_{0}$, $N$ for both case I ($n_{0}$ = $442\pm21$, $N$ = $749\pm27$) and case II ($n_{0}$ = $510\pm22$, $N$ = $817\pm29$), respectively.  For comparison purposes, we also include results for samples with $P_{mem} \geq 0.2$ that have values for case I ($n_{0}$ = $114\pm11$, $N$ = $194\pm14$) and case II ($n_{0}$ = $166\pm13$, $N$ = $246\pm16$), respectively.

Figure \ref{Fig19} shows the observed and estimated number of stars as a function of flare occurrences.  The observed distribution seems to follow a Poisson-like behavior, but both cases do not represent the values of $n_{k}$ where $k > 3$.  \citet{amb70} suggested that a small group of stars with high flare rates could explain the discrepancy between observed flare stars and those predicted by Equation 18.  We agree with this suggestion because individual flare stars may have time-dependent flare rate caused by activity cycle or variations in active regions, as on the Sun (e.g., \citealt{lac76, hil10, hil11}). 

\begin{deluxetable}{cccccc}
\tabletypesize{\footnotesize}
\tablecaption{Power-law slopes of flare frequency distributions \label{Tab5}}
\tablewidth{0pt}
\tablehead{
\colhead{$r_{0}$} & \colhead{$c_\mathrm{lower}$} & \colhead{$c_\mathrm{upper}$} & \colhead{$\beta$} & \colhead{$\log E_\mathrm{min}$} & \colhead{$\log E_\mathrm{max}$}
}
\startdata
17--18\phn &  $>$17.83 ($>$14.39)\phn &  $>$18.77 ($>$15.30)\phn & $>$$0.62\pm0.05$ ($>$0.52$\pm$0.03)\phn &32.90 (32.90)\phn & $>$35.23  ($>$35.23)\phn\\
18--19\phn &  26.01 (26.61)\phn &  27.02 (27.60)\phn & $0.87\pm0.05$ $(0.89\pm0.06)$\phn & 32.90 (32.90)\phn & 34.45 (34.39)\phn\\           
19--20\phn &  30.37 (27.71)\phn &  31.34 (28.65)\phn & $1.01\pm0.05$ $(0.93\pm0.08)$\phn & 32.90 (32.90)\phn & 34.18 (34.18)\phn\\          
20--21\phn &  30.22 (29.44)\phn &  31.01 (30.20)\phn & $1.00\pm0.05$ $(0.97\pm0.03)$\phn & 32.90 (32.90)\phn & 34.31 (34.31)\phn\\           
21--22\phn &  36.98 (17.57)\phn &  37.79 (18.36)\phn & $1.21\pm0.05$ $(0.62\pm0.02)$\phn & 32.90 (32.90)\phn & 34.48 (34.48)\phn\\           
22--23\phn &  32.34 (\nodata)\phn &  33.21 (\nodata)\phn & $1.08\pm0.04$ (\nodata)\phn & 32.90 (\nodata)\phn & 34.14 (\nodata)\phn
\enddata
\tablecomments{The values in the parentheses are for stars with membership information ($P_{mem} \geq 0.2$).}
\end{deluxetable}

\subsubsection{Flare frequency distribution}
We use a cumulative energy distribution of flare frequency (also known as flare frequency distribution; FFD).  This method is the most widely used to estimate the frequency with which a flare of a particular energy is seen (e.g., \citealt{lac76,ger83,ish91,hil11}).  The cumulative frequency $\nu (E)$ at energy $E$ is defined as the number of flares with energy greater than $E$ per unit time (normalized by the number of \emph{real} flare stars $N_{\star}$), where $\nu (E)$ = $N^{cum} (>E)/\tau_{obs} \cdot N_{\star}$.  If the differential frequency distribution obeys a power-law relation with slope $\alpha$ and a cutoff energy at $E_{max}$, i.e., $dN / dE \propto E^{-\alpha}$, the cumulative frequency distribution can be estimated by a linear fit with a slope of $\alpha - 1 (=\beta)$:
\begin{equation}
\log \nu (E) = c - \beta \log E,
\end{equation} where $c$ is a constant.  The uncertainties in the slope estimates are taken to be $\beta / \sqrt(N_{\star})$.  Since there is no way to estimate the contamination of field flare stars in the cluster sample, we set the lower ($c_{lower}$; $N_{\star} \simeq N_{s}$) and upper ($c_{upper}$; $N_{\star} \simeq N_{f}$) limits for the cumulative flare frequency.  For the cluster flare stars, the real flare frequency will similar to or less than the latter limit.

\begin{figure}[!t]
  \centering
  \includegraphics[width=1.0\linewidth, angle=0]{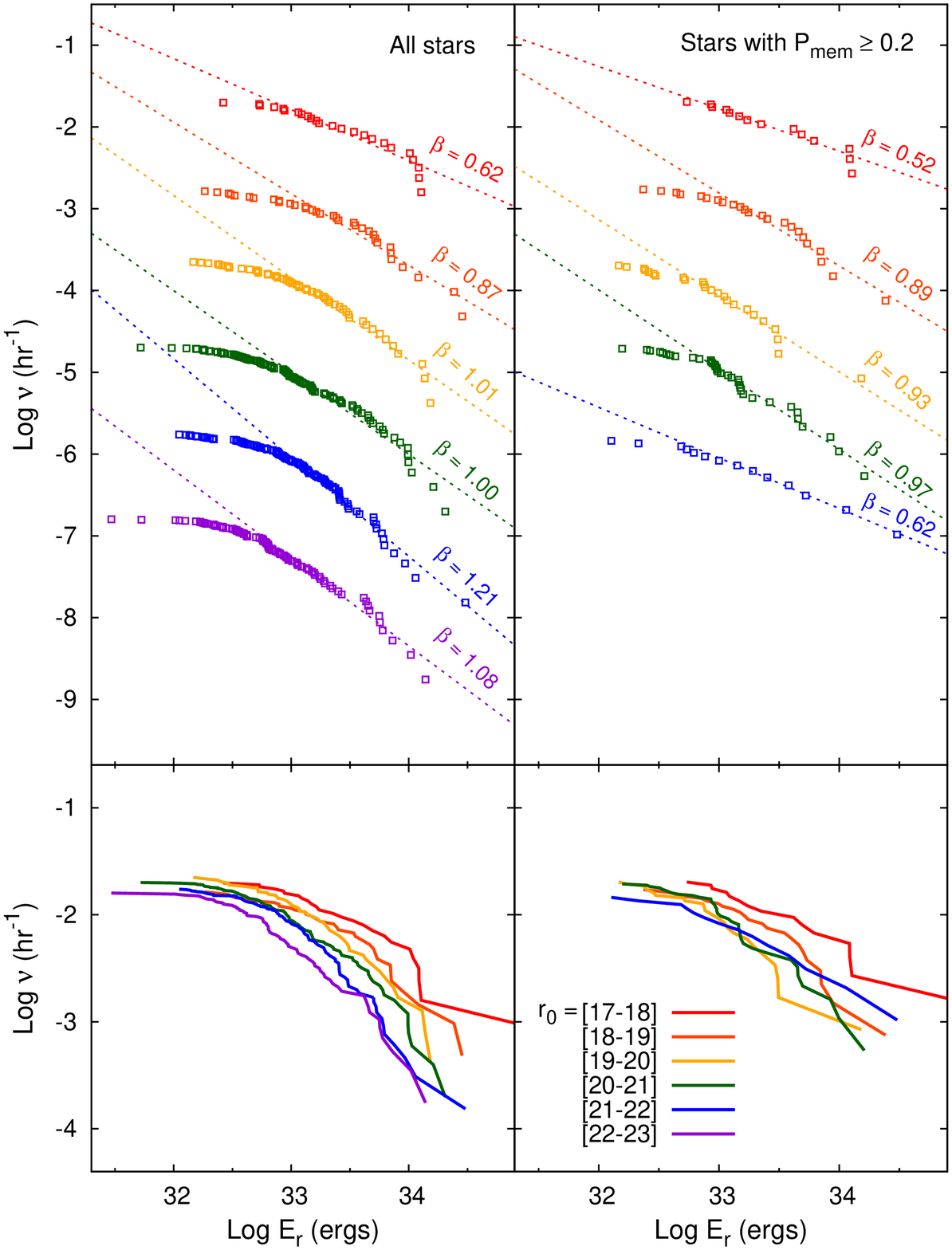}
  \caption{Upper limit for cumulative flare frequency distributions $\nu(E)$ vs. $r$-band flare energy (vertically shifted for each magnitude bin, in order of stellar magnitude down to $r$=22--23).  The best fits to a power-law model in Table \ref{Tab5} are shown as dashed lines for the adopted range of flare energies.  For comparison, we also show the individual FFDs with no shift at the bottom panels.  As expected, strong flares are much less frequent than weak ones.}
  \label{Fig20}
\end{figure}

We consider the completeness of our sample when estimating the power-law index for flare frequency.  As shown in the Figure \ref{Fig12}, observed distribution of flare energies exhibits a turn-over near $\log E_{r} = 32.9$.  It is certain that we do not miss the flares with energies above this completeness limit.  The slopes are obtained from a least-squares linear regression to the flares with this cut-off energy (Table \ref{Tab5}).  However, the low-energy flares below this cut-off are not small enough to be missed observationally (see Figure \ref{Fig14} and Figure \ref{Fig16}).   This issue will be addressed in a separate work where we use a thorough Monte Carlo simulations to examine observational incompleteness and also to seek possible real break in the FFD of flare energies.

In the top panels of Figure \ref{Fig20}, we show the best-fit power-law slopes of the FFDs for each magnitude bin with different colors.  The upper limits of flare rate are plotted against the flare energy for all cluster sample (left panels) and samples with $P_{mem} \geq 0.2$ (right panels), respectively.  Our FFDs follow the straight-line power-law form only for the higher energies, but these resemble a log-normal-like distribution.  The low-energy turnover seen at all magnitude bin, while at the high-energy end we can see a gradual change in the shape of the FFDs due to finite total observing time.  Our cluster sample shows that the power-law slopes of FFDs range between $-0.62$ and $-1.21$ for the adopted energy range.  For stars with $P_{mem} \geq 0.2$, these values are similar within the uncertainties but somewhat smaller than the former case.  The slopes are slightly steeper for faint stars ($\beta \sim 1.0$) compared to bright ones ($\beta < 1.0$).  However, we caution that the latter groups may have suffered from the relatively small sample size.  For comparison purposes, we also show the individual FFDs with no shift (bottom panels of Figure \ref{Fig20}).  The average shape of FFDs is very similar within the uncertainties, showing that the frequency of flares decreases as a function of increasing flare energy.  Strong flares are observed $\sim$10--100 times less frequently than weak flares.  As expected, the strong flares on the brighter stars are more frequent than fainter ones.

\begin{figure}[!t]
  \centering
  \includegraphics[width=1.0\linewidth, angle=0]{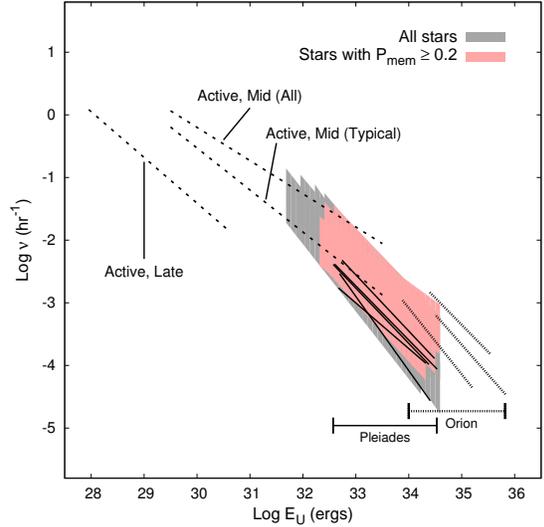}
  \caption{Cumulative flare frequency vs. $U$-band flare energy for this work and those for other statistical studies, taken from \citet{hil11} and \citet{ger83} for flare stars in the Solar vicinity and in the Orion and Pleiades clusters, respectively.  In the case of mid-M type flare stars, one more line is added to represent the typical FFD for these spectral types (excluding the stars with exceptionally high flare rates such as AD Leo, YZ CMi, EV Lac, and EQ Peg).  The two clusters contain a sample of late K- and early M-dwarfs and early to mid M-dwarfs, respectively.  The slopes of their FFDs cover a large range of flare energies from $E_{U}$=10$^{28}$ to $E_{U}$=10$^{36}$ erg.  Our results are indicated by the shaded region which are determined by the lower and upper limits of individual FFD values, except for the first row in Table \ref{Tab5}.  We find a good agreement with the overall trend.}
  \label{Fig21}
\end{figure}

Figure \ref{Fig21} shows a comparison between the FFDs in this work and those by other statistical studies, particularly for open clusters and field stars in the solar neighborhood.  The first comparison data set is a sample of nearby young clusters known to contain sizable flare stars: Pleiades and Orion.  Each power-law slopes for stars of similar brightness was taken from the figure of \citet{ger83}.  The second comparison sample is taken from the \citet{hil11} in which we only used the FFDs for active, mid-M dwarfs (M3--M5) and active, late-M dwarfs (M6--M8).  The gradient of FFD is a critical factor in understanding the energy dependence of the flare frequency.  The FFDs of these samples are approximately a power law $\nu(E) \sim E^{-\beta}$ with $\beta$ $\sim$ 0.5--1.2, and their power-law slopes cover a large range of flare energies from $E_{U}$=10$^{28}$ to $E_{U}$=10$^{36}$ erg.  Our results follow a general picture of FFD in the same energy range with $\beta$ $\sim$ 0.6--1.2 for all stars (black-shaded areas) and $\beta$ $\sim$ 0.5--1.0 for stars with $P_{mem} \geq 0.2$ (red-shaded areas).
 
\section{Summary}
We present the statistical properties of flare variability as a direct evidence for stellar activity in low-mass stars.  Our study is a rare attempt to estimate flare rates and physical properties among many stars of the same age and mass group, and the main results are as follows: 

We monitored light variations of nearly 15,800 red dwarfs in the M37 cluster field, and then successfully identified 420 flare events from 312 cluster sample and 184 flare events from 107 field sample, respectively.  Among the cluster sample, only 82 stars have cluster membership probability with $P_{mem} \geq 0.2$.  Most of flare stars fall close to the sequence of the cluster as expected, while the remaining stars lie within about 200 pc from the Galactic plane (i.e., young, active K--M dwarfs in the thin disk).  These red dwarfs produce serendipitous flares which are energetic enough to be observed even in the $r$-band.  The temporal and morphological characteristics of flare light curves are almost the same as those found in $U$-band observations.  For the cluster sample, we also found many large-amplitude flares with inferred $\Delta u > 6$ mag.  These rare events are of considerable interest due to the influence of their powerful radiation on space weather and planetary habitability.

We find that statistically significant correlations exist between the flare energies and other key parameters.  Flare energy is tightly correlated with flare duration and peak luminosity (i.e., fractional area coverage) in log-log space, regardless of stellar magnitude.  For the group of stars with similar brightness, the flare frequency distributions can be approximated by a power-law form $\nu(E) \sim E^{-\beta}$ with $\beta$ $\sim$ 0.62--1.21 for all flare stars and $\beta$ $\sim$ 0.52--0.97 for stars with membership information ($P_{mem} \geq 0.2$), which are in agreement with previous studies on other open clusters and solar neighborhood stars.  These results suggest that stellar flares are likely powered by similar physical mechanisms that initiate and drive the flaring event.  Moreover, flare stars in young- and intermediate-aged open clusters produce up to a thousand times more flare energy than those of field stars in the solar neighborhood.  Our data of M37 falls nicely between the young open clusters and nearby field stars.

\acknowledgments
This research was supported by Basic Science Research Program through the National Research Foundation of Korea (NRF grant 2011-0030875).  Y.-I.B. acknowledges the support from KASI-Yonsei DRC program of Korea Research Council of Fundamental Science and Technology (DRC-12-2-KASI).  We thank J. R. A. Davenport for his assistance with the two-component flare model.


\clearpage

\renewcommand{\thetable}{A\arabic{table}}
\renewcommand{\theequation}{A\arabic{equation}}
\renewcommand\thefigure{A\arabic{figure}}   
\setcounter{equation}{0}  
\setcounter{figure}{0}
\setcounter{table}{0}

\begin{deluxetable}{ccccccccccccccccc}
\tabletypesize{\tiny} 
\rotate
\tablecaption{Observational parameters of 604 flare events detected in the whole sample \label{TabA1}}
\tablewidth{0pt}
\tablehead{
\colhead{} & \colhead{} & \colhead{} & \colhead{} & \colhead{} & \colhead{} & \multicolumn{2}{c}{$\tau_{0.9}$} & \colhead{} & \multicolumn{2}{c}{$\tau_{0.5}$} & \colhead{} & \multicolumn{2}{c}{$\tau_{0.2}$} & \colhead{} & \multicolumn{2}{c}{$\tau_{0}$}\\
\cline{7-8} \cline{10-11} \cline{13-14} \cline{16-17} \\
\colhead{} & \colhead{$r$} & \colhead{} & \colhead{$t_{peak}$} & \colhead{$\Delta m_\mathrm{peak}$} & \colhead{$\Delta m_\mathrm{base}$} & \colhead{$t_\mathrm{0.9,rise}$} & \colhead{$t_\mathrm{0.9,decay}$} & \colhead{} & \colhead{$t_\mathrm{0.5,rise}$} & \colhead{$t_\mathrm{0.5,decay}$} & \colhead{} & \colhead{$t_\mathrm{0.2,rise}$} & \colhead{$t_\mathrm{0.2,decay}$} & \colhead{} & \colhead{$t_\mathrm{0,rise}$} & \colhead{$t_\mathrm{0,decay}$}\\
\colhead{VarID} & \colhead{(mag)} & \colhead{FlareID} & \colhead{(days)} & \colhead{(mag)} &  \colhead{(mag)} & \colhead{(min)} & \colhead{(min)} & \colhead{} & \colhead{(min)} & \colhead{(min)} & \colhead{} & \colhead{(min)} & \colhead{(min)} & \colhead{} & \colhead{(min)} & \colhead{(min)}
}
\startdata
V42\phn & 20.690\phn & F1\phn & 53742.40307\phn & 0.096\phn & \nodata\phn & 0.25\phn & 1.25\phn & & 1.54\phn & 4.15\phn & & 3.46\phn & 6.71\phn & & 5.02\phn & 19.67\phn\\
\phn & \phn & F2\phn & 53742.46247\phn & 0.137\phn & 0.051\phn & 0.34\phn & 4.57\phn & & 1.69\phn & 13.19\phn & & 4.68\phn & \nodata & & 12.46\phn & \nodata\\
V50\phn & 19.990\phn & F1\phn & 53730.34740\phn & 0.221\phn & \nodata\phn & 0.90\phn & 0.39\phn & & 4.20\phn & 2.05\phn & & 9.67\phn & 23.53\phn & & \nodata & 111.55\phn\\
\phn & \phn & F2\phn & 53733.19093\phn & 0.176\phn & \nodata\phn & 0.10\phn & 0.27\phn & & 0.52\phn & 1.69\phn & & 0.83\phn & 20.18\phn & & 1.04\phn & 66.29\phn\\
\phn & \phn & F3\phn & 53735.19767\phn & 0.035\phn & \nodata\phn & 3.13\phn & 1.11\phn & & 4.52\phn & 5.45\phn & & 5.13\phn & 14.17\phn & & 5.53\phn & 17.70\phn\\
\phn & \phn & F4\phn & 53737.33376\phn & 0.187\phn & \nodata\phn & 1.46\phn & 0.56\phn & & 8.87\phn & 2.70\phn & & 9.73\phn & 12.10\phn & & 13.72\phn & 108.37\phn\\
V69\phn & 20.809\phn & F1\phn & 53732.22482\phn & 0.061\phn & \nodata\phn & 0.13\phn & 0.78\phn & & 0.64\phn & 3.44\phn & & 1.02\phn & 9.58\phn & & 1.46\phn & 17.36\phn\\
\phn & \phn & F2\phn & 53732.35513\phn & 0.133\phn & \nodata\phn & \nodata & 1.06\phn & & \nodata & 15.86\phn & & \nodata & 56.99\phn & & \nodata & 82.70\phn\\
\phn & \phn & F3\phn & 53742.35427\phn & 0.281\phn & \nodata\phn & 0.55\phn & 0.91\phn & & 2.17\phn & 5.49\phn & & 2.90\phn & 11.78\phn & & \nodata & 96.72\phn\\
\enddata
\tablecomments{Table \ref{TabA1} is published in its entirety in the electronic edition of {\it Astrophysical Journal}.  A portion is shown here for guidance regarding its form and content.  The identification number (VarID) is a number uniquely identifying each flare star (notations are taken from PaperII).  For each star, the identification number (FlareID) of each flare event is listed in time order.}
\end{deluxetable}
\clearpage

\begin{deluxetable}{cccccccccc}
\tabletypesize{\footnotesize}
\tablecaption{Physical parameters of 420 flare events detected in the cluster sample\label{TabA2}}
\tablewidth{0pt}
\tablehead{
\colhead{} & \colhead{} & \colhead{} & \colhead{} & \colhead{} & \colhead{Quiet} & \colhead{Peak} & \colhead{} & \colhead{}\\
\cline{6-7} \\
\colhead{} & \colhead{$r$} & \colhead{} & \colhead{$\tau_{0}$} & \colhead{$P_{r}$} & \colhead{log $L_\mathrm{r}$} & \colhead{log $L_\mathrm{r, peak}$} & \colhead{log $E_{r}$} & \colhead{$P_{mem}$\tablenotemark{a}}\\
\colhead{VarID} & \colhead{(mag)} & \colhead{FlareID} & \colhead{(minute)} & \colhead{(second)} & \colhead{(erg s$^{-1}$)} & \colhead{(erg s$^{-1}$)} & \colhead{(erg)} & \colhead{(\%)}
}
\startdata
V78\phn & 20.017\phn & F1\phn & $>$12.92\phn & $>$12\phn & 31.288\phn & $>$29.934\phn & $>$32.377\phn & 0.288\phn \\
V86\phn & 18.086\phn & F1\phn & 76.88\phn & 61\phn & 32.061\phn & 30.643\phn & 33.850\phn & 0.105\phn \\
\phn & \phn & F2\phn & 35.42\phn & 23\phn & \phn & 30.430\phn & 33.428\phn & \phn\\
V91\phn & 20.553\phn & F1\phn & 58.06\phn & 44\phn & 31.074\phn & 29.916\phn & 32.719\phn & 0.119\phn \\
\phn & \phn & F2\phn & $>$67.49\phn & $>$126\phn & \phn & $>$29.954\phn & $>$33.175\phn & \phn \\
V112\phn & 22.181\phn & F1\phn & 83.03\phn & 734\phn & 30.423\phn & 30.010\phn & 33.289\phn & 0.041\phn \\
V144\phn & 20.899\phn & F1\phn & 101.83\phn & 696\phn & 30.936\phn & 30.673\phn & 33.779\phn & 0.050\phn \\
\phn & \phn & F2\phn & 122.31\phn & 547\phn & \phn & 30.294\phn & 33.674\phn & \phn\\
V169\phn & 20.835\phn & F1\phn & 123.50\phn & 145\phn & 30.961\phn & 30.117\phn & 33.124\phn & 0.000\phn \\
V191\phn & 22.002\phn & F1\phn & 27.92\phn & 819\phn & 30.495\phn & 30.819\phn & 33.408\phn & 0.004\phn \\
\enddata
\tablecomments{Table \ref{TabA2} is published in its entirety in the electronic edition of {\it Astrophysical Journal}.  A portion is shown here for guidance regarding its form and content.}
\tablenotetext{a}{Cluster membership probabilities are taken from \citet{nun15}.}
\end{deluxetable}

\clearpage
\end{document}